\documentclass{aa}  

\usepackage{natbib}
\bibpunct{(}{)}{;}{a}{}{,}
\usepackage[T1]{fontenc}
\usepackage{ae,aecompl}
\usepackage{adjustbox}

\usepackage{multicol}        
\usepackage{multirow}        

\usepackage{epsfig,amstext,floatfig,alltt,fancyhdr,setspace}
\usepackage{graphicx}
\usepackage{lmodern} 
\usepackage{color}
\usepackage{ucs}
\usepackage{url}
\usepackage{mathrsfs}
\usepackage{amsmath} 
\usepackage{amssymb}
\usepackage{appendix}
\usepackage{varioref}
\usepackage{txfonts}
\usepackage{enumerate}
\usepackage{magazin_abbrev}
\definecolor{purple}{RGB}{76, 0,153}

\def \Eqt{Eq.\;}
\def \sect{Sect.\thinspace}
\def \fig{Fig.\thinspace}
\def \tab{Table\thinspace}
\def \App{Appendix\thinspace}

\def\Om{{\Omega_{\rm m}}}

\def\Cov{\mathbb{C}}

\def\d{{\rm d}}

\newcommand{\mcal}{{\sc metacalibration }}
\newcommand{\im}{{\sc im3shape }}

\newcommand{\hmcode}{{\sc HMCode }}

\begin{document} 



\title{KiDS+VIKING-450 and DES-Y1 combined: Mitigating baryon feedback uncertainty with COSEBIs}

\author{Marika Asgari\inst{1}\fnmsep\thanks{E-mail: ma@roe.ac.uk}
		\and
		Tilman Tröster\inst{1}
		\and
		Catherine Heymans\inst{1,2}
		\and
		Hendrik Hildebrandt\inst{2}
		\and
		Jan Luca van den Busch\inst{2}
		\and
		Angus H. Wright\inst{2}
		\and
		Ami Choi\inst{3}
		\and
		Thomas Erben\inst{4}
		\and
		Benjamin Joachimi\inst{5}
		\and
		Shahab Joudaki\inst{6}
		\and
		Arun Kannawadi\inst{7}
		\and
		Konrad Kuijken\inst{7}
		\and
		Chieh-An Lin\inst{1}
		\and
		Peter Schneider\inst{4}
		\and
		Joe Zuntz\inst{1} 
          }
   \institute{Institute for Astronomy, University of Edinburgh, Royal Observatory,
Blackford Hill, Edinburgh, \; EH9 3HJ, U.K.
         \and
         German Centre for Cosmological Lensing, Astronomisches Institut, Ruhr-Universität Bochum, Universitätsstr. 150, 44801, Bochum, Germany
         \and
         Department of Physics and Astronomy, University College London, Gower Street, London WC1E 6BT, UK
         \and
         Argelander-Institut für Astronomie, Auf dem Hügel 71, 53121 Bonn, Germany
         \and
         Center for Cosmology and AstroParticle Physics, The Ohio State University, 191 West Woodruff Avenue, Columbus, OH 43210, USA
	    \and
Department of Physics, University of Oxford, Denys Wilkinson Building, Keble Road, Oxford OX1 3RH, UK
         \and
         Leiden Observatory, Leiden University, Niels Bohrweg 2, 2333 CA Leiden, The Netherlands
             }

\date{Received XXX; accepted YYY}

\abstract{We present cosmological constraints from a joint cosmic shear analysis of the Kilo-Degree Survey (KV450) and the Dark Energy Survey (DES-Y1), conducted using Complete Orthogonal Sets of E/B-Integrals (COSEBIs). With COSEBIs we isolate any B-modes which have a non-cosmic shear origin and demonstrate the robustness of our cosmological E-mode analysis as no significant B-modes are detected.  We highlight how COSEBIs are fairly insensitive to the amplitude of the non-linear matter power spectrum at high $k$-scales, mitigating the uncertain impact of baryon feedback in our analysis. COSEBIs, therefore,  allow us to utilise additional small-scale information, improving the DES-Y1 joint constraints on $S_8=\sigma_8(\Om/0.3)^{0.5}$ and $\Om$ by $20\%$. Adopting a flat $\Lambda$CDM model we find $S_8=0.755^{+0.019}_{-0.021}$, which is in $3.2\sigma$ tension with the Planck Legacy analysis of the cosmic microwave background.}


\titlerunning{KV450 and DES-Y1 with COSEBIs}
\authorrunning{M. Asgari et al.}
\maketitle


\section{Introduction}

\label{sec:introduction}

Cosmic shear is one of the primary probes for many current (KiDS\footnote{Kilo Degree survey: http://kids.strw.leidenuniv.nl/}, DES,\footnote{Dark Energy Survey: www.darkenergysurvey.org} and HSC\footnote{Hyper-Suprime Camera Survey: hsc-release.mtk.nao.ac.jp}) and future (Euclid\footnote{Euclid mission: www.euclid-ec.org}, LSST\footnote{The Large Synoptic Survey Telescope: www.lsst.org} and WFIRST\footnote{Wide Field Infrared Survey Telescope: www.nasa.gov/wfirst})  cosmological surveys, 
since gravitational lensing probes the matter distribution directly and thus does not require the use of biased tracers such as galaxy positions \citep{Kilbinger_review}.
Cosmic shear analysis is, however, not without its own challenges. For example, inaccuracies in the data processing pipeline can result in systematic biases that mimic the signal \citep[e.g.][]{hoekstra:2004}.  There are also challenges on the theoretical side of the analysis, such as the intrinsic alignment of galaxies which have a similar shape to a cosmic shear signal \citep[e.g.][]{joachimi/etal:2015}
and the modelling of small scales which depend on  baryon feedback processes \citep[e.g.][]{semboloni/etal:2011}. In addition, the redshift distribution of galaxies needs to be calibrated accurately to avoid systematic shifts in the inferred cosmological parameters \citep[e.g.][]{vanwaerbeke/etal:2006}.

One of the primary robustness tests for cosmic shear studies is to check that the cosmic shear data contain no significant B-modes.
Gravitational lensing only produces E-modes, while unaccounted systematics in the data can produce both E and B-modes. 
We can therefore use B-modes to assess the quality of the data. 
The primary statistics used in the cosmic shear analyses of KiDS \citep[Kilo Degree Survey;][]{hildebrandt/etal:2017,hildebrandt/etal:2019}, 
DES \citep[Dark Energy Survey;][]{Becker15,troxel/etal:2018a}, DLS \citep[Deep Lens Survey;][]{Jee2016}, 
and CFHTLenS \citep[Canada France Hawaii Telescope Lensing Survey;][]{heymans/etal:2013} 
are the shear two-point correlation functions (2PCFs), $\xi_\pm$ \citep{Kaiser92}.
These two-point statistics mix E/B-modes and are therefore ill-suited for detecting systematics that introduce B-modes. 
In this paper we will advocate the use of COSEBIs \citep[Complete Orthogonal Sets of E/B-Integrals;][]{SEK10} which are designed to cleanly separate E/B-modes from $\xi_\pm$ measured on a finite angular range, while maintaining all E/B-mode information,
and show how we can use them to mitigate the effects of baryon feedback in the analysis\footnote{Another way to reduce the effect of baryons is to directly estimate power spectra from the data  \citep[For example k-cut power spectra,][]{taylor/etal:2018}. Power spectrum estimators however suffer from measurement biases that do not affect COSEBIs \citep[see][for a more detailed discussion]{asgari/etal:2019a}.}.
Similar to $\xi_\pm$, COSEBIs are unaffected by masking, as long as the masks are uncorrelated with the matter distribution\footnote{This is not true for their covariance (see \App\ref{app:covariance}).}.
This simplifies the analysis compared to other estimators, such as pseudo-$C_\ell$ methods \citep[see, for example,][]{Hikage11,asgari_pcls}, which were used as the primary statistic in HSC \citep{hikage/etal:2018}. Other methods that focus on finding a minimal set of nuisance parameters using a principal component analysis \citep{eifler/etal:2015} can also be applied to mitigate the effect of baryon feedback \citep{huang/etal:2019}. 

COSEBIs were used in non-tomographic analyses of CFHTLenS \citep{kilbinger/etal:2013} and SDSS \citep[Sloan Digital Sky Survey;][]{huff/etal:2014}.
In this work we present the first tomographic analysis of cosmic shear data with COSEBIs, and place constraints on cosmological parameters. 
We measure COSEBIs with the first 450 deg$^2$ of data from KiDS including overlapping data from the VISTA infrared galaxy surveys \citep[KiDS+VIKING-450: KV450;][]{wright/etal:2019} as well as the first year of data from the Dark Energy Survey \citep[DES-Y1;][]{DrlicaWagner/etal:2018, Zuntz/etal:2018}.

Although COSEBIs were originally designed to completely separate E/B-modes, they have many more advantages over other statistics used in cosmic shear analysis, in particular 2PCFs.
While $\xi_\pm$ are continuous functions of the angular scale, $\theta$, for analysis and measurement purposes $\xi_\pm$ are binned in $\theta$ to reduce the number of data points. 
The binning of 2PCFs can affect both the theoretical predictions \citep[][]{troxel/etal:2018b,asgari/etal:2019a} and the covariance of $\xi_\pm$, which has not been accounted for so far.
COSEBIs have discrete and well-defined modes, and therefore do not suffer from these complications. 
In addition, the first few modes of COSEBIs are sufficient for capturing the full information in a cosmological analysis.
They therefore provide a natural form of data compression per pair of redshift bins \citep{Asgari12}.
When the data are divided into many redshift bins, complementary data compression methods can also be applied \citep[see for example,][]{AS2014}.

Given a fixed angular range, COSEBIs are significantly less sensitive to small scale physical effects compared to $\xi_\pm$, as we will show in this paper. These scales are affected by baryon feedback processes, which are challenging to model, since they depend on our understanding and ability to simulate complicated physics.
Hydrodynamical simulations have been used in conjunction with a halo-model analytical approach to account for the effect of baryon feedback on the matter power spectrum \citep[]{mead/etal:2015}; however, these simulations give differing results based on their underlying assumptions and subgrid physics \citep{chisari/etal:2018}. 
As a result, the theoretical predictions for scales affected by baryon feedback are less reliable. 
For cosmic shear analyses baryon feedback has so far either been modelled and marginalised over using the halo model \citep[H20 hereafter]{hildebrandt/etal:2020} or single parameter fits to simulations \citep{hikage/etal:2018} or scale cuts have been applied to the measurements to minimise its effects \citep[][T18 hereafter]{troxel/etal:2018a}. 
In this paper, we show that COSEBIs are significantly less sensitive to baryon feedback in comparison to 2PCFs. This allows us to confidently include small angular scale information, previously excluded from the T18 analysis of the DES-Y1 data, resulting in improved constraints. 

Cosmic shear data from both KV450 and DES-Y1 are publicly available. 
These datasets have no overlapping regions, such that the cross-covariance between the surveys can be neglected, allowing for a simple joint analysis of these two surveys. 
The cosmic shear analyses of both of these surveys yielded consistent, but smaller, $S_8\equiv\sigma_8(\Om/0.3)^{0.5}$ values than the Planck Legacy analysis of the cosmic microwave background \citep{Planck2018}.  It is therefore interesting to ask whether any 
tension between these two probes increases if we combine KV450 and DES-Y1. 
Recently, \cite{joudaki/etal:2019} carried out a joint analysis of KV450 and DES-Y1 using 2PCFs. 
They incorporated a spectroscopic calibration for the redshift distributions of galaxies and a consistent set of priors. 
\cite{joudaki/etal:2019} found an $S_8$ value that is in tension with Planck by $2.5\sigma$.
Here we follow the same procedure for combining these surveys, using COSEBIs instead of 2PCFs.  Our fiducial analysis also adopts the alternative spectroscopic calibration approach for the DES-Y1 redshift distributions.

This paper is organised as follows: \sect\ref{sec:method} introduces 2PCFs and COSEBIs as well as a comparison between these two statistics. In \sect\ref{sec:data}, we briefly present the KV450 and DES-Y1 data and our analysis pipeline. In \sect\ref{sec:results} we show the results of the cosmological analyses of these datasets and their combination. We conclude in \sect\ref{sec:conclusions}. Details of the analytical covariance matrix of COSEBIs is described in \App\ref{app:covariance}. \App\ref{app:triangle} shows the constraints on the full sets of cosmological parameters that we have sampled.

\section{Methods}

\label{sec:method}

In this paper we focus on the use of COSEBIs in cosmic shear analysis and compare it to shear two-point correlation functions (2PCFs), which have been the primary statistics used in most cosmic shear surveys to date. 
In the following sections we briefly define 2PCFs and COSEBIs, their estimation methods, and theoretical modelling. 
We then compare the sensitivity of 2PCFs and COSEBIs to different Fourier scales.
The covariance of COSEBIs is derived in \App\ref{app:covariance}.

\subsection{Shear two-point correlation functions}

Shear two-point correlation functions are defined as the correlation between the shear estimates of two galaxies at a given angular separation, $\theta$, on the sky.
The shear 2PCFs commonly employed in cosmic shear analysis are defined as
\begin{align}
\label{eq:xi-pm-def}
\xi_\pm(\theta) & =\langle\gamma_\mathrm{t} \gamma_\mathrm{t} \rangle (\theta)\pm 
\langle\gamma_\mathrm{\times} \gamma_\mathrm{\times} \rangle (\theta) \ ,
\end{align} 
where $\gamma_\mathrm{t}$ is the tangential and $\gamma_\mathrm{\times}$ is the cross-component of the shear { and the angular brackets denote an ensemble average over all pairs of galaxies with an angular distance equal to $\theta$.} 
Galaxies have an intrinsic ellipticity associated with them, which is distorted as a result of gravitational lensing by intervening matter. 
An unbiased and noise-free measurement of the observed ellipticity is therefore equal to a combination of their shear and intrinsic ellipticity. The estimator for the shear 2PCFs in \Eqt\eqref{eq:xi-pm-def} is given by
\begin{equation}
\label{eq:xipm_meaure}
\hat\xi_\pm(\theta)=\frac{\sum_{ab} w_a w_b
 \left[\epsilon_{\rm t}({\pmb x}_a)\epsilon_{\rm t}({\pmb x}_b)
 \pm\epsilon_{\times}({\pmb x}_a)\epsilon_{\times}({\pmb x}_b)\right]}
 {\sum_{ab} w_a w_b (1+m_a)(1+m_b)} \ ,
\end{equation}
where $\epsilon_{\rm t}({\pmb x}_a)$ and $\epsilon_{\times}({\pmb x}_a)$ are the tangential and cross components of the measured ellipticity for a galaxy at position ${\pmb x}_a$ { defined with respect to the line connecting the pair of galaxies at ${\pmb x}_a$ and ${\pmb x}_b$}  . The sum extends over all galaxies with $|{\pmb x}_a-{\pmb x}_b|$ within the angular bin, $\theta$. In \Eqt\eqref{eq:xipm_meaure}, $w$ and $m$ are the weight and multiplicative bias associated with each galaxy, respectively. The multiplicative bias correction is applied here because, in practice, measuring the shapes of galaxies in the presence of noise results in a biased estimate \citep{melchior_viola:2012}.
This bias can be calibrated using image simulations, as it was for the KV450 data \citep{kannawadi/etal:2019} or from the dataset itself using \mcal \citep[see][]{sheldon_huff:2017}. In the latter case the weights are all set to unity and the factors of $(1+m)$ in \Eqt\eqref{eq:xipm_meaure} are replaced by the response of the measurement method to shear.  
Note that even in the case of \mcal\!\!, the response correction may not correct for all of the systematics in the shear measurement and therefore an extra $m$-bias correction may have to be applied \citep{Zuntz/etal:2018}. 
The multiplicative bias correction can also be applied to an ensemble of galaxies, once their binning and weighting is applied, as a single $m$-bias value per galaxy grouping (for example per redshift bin).

The expectation of the shear 2PCFs can be derived using their relation to the 2D convergence power spectrum, $P_\kappa(\ell)$:
\begin{align}
 \label{eq:xipm}
& \xi_\pm(\theta)=\int_0^\infty \frac{\mathrm{d}\ell\:\ell}{2\pi}\:
 \mathrm{J}_{0/4}(\ell\theta)\:P_\kappa(\ell)\ ,
\end{align} 
where  $\mathrm{J}_{0/4}$ are the zeroth and fourth-order Bessel functions of the first kind. Assuming a modified Limber approximation \citep{LoverdeAfshordi08,kilbinger/etal:2017} we can relate $P_\kappa(\ell)$ to the 3D matter power spectrum, $P_\delta$, for redshift bins $i$ and $j$,
\begin{align}
\label{eq:limber}
 P_\kappa^{ij}(\ell)= \frac{9H_0^4\Omega_\mathrm{m}^2}{4c^4}
 \int_0^{\chi_\mathrm{h}}\mathrm{d}\chi\:\frac{g^i(\chi)g^j(\chi)}{a^2}\:P_\delta
 \left(\frac{\ell+1/2}{f_\mathrm{K}(\chi)},\chi\right)\;,
\end{align} 
where $H_0$ is the Hubble parameter, $\Om$ is the matter density parameter, $c$ is the speed of light, $\chi$ is the comoving distance, $\chi_\mathrm{h}$ is the comoving distance to the horizon, $f_\mathrm{K}(\chi)$ is the comoving angular diameter distance, and
\begin{equation}
g^i(\chi) = \int_{\chi}^{\chi_\mathrm{h}} \mathrm{d}\chi'\: p^i_{\chi}(\chi')\:
\frac{f_\mathrm{K}(\chi'-\chi)}{f_\mathrm{K}(\chi')}\;,
\label{eq:gchi}
\end{equation} 
where $p^i_{\chi}$  is the probability density of sources in comoving distance for redshift bin $i$. 

The observed ellipticity field does not correspond to a physical convergence field in general (when curl modes are present). 
We therefore expect to have both E and B-mode power spectra in the field. 
In the absence of systematics in the data, however, we expect to find B-modes to be consistent with zero\footnote{Effects such as source clustering \citep{schneider/etal:2002b}, contributions beyond Born approximation \citep{Schneider98}, exotic cosmological models \citep[see for example][]{thomas/etal:2017} and tidal intrinsic alignment models \citep[e.g.][]{blazek/etal:2015} can produce B-modes. They are of a level, however, that can currently be neglected.},
allowing B-modes to be used to assess the credibility of the data 
\citep[see for example][]{hoekstra:2004, kilbinger/etal:2013, asgari/etal:2017,asgari/etal:2019a,asgari_heymans:2019}. 
Shear 2PCFs mix E and B-modes:
\begin{align}
\label{eq:xipmPower}
 \xi_\pm(\theta) &=\int_0^{\infty} \frac{\d \ell\, \ell}{2\pi} \rm{J}_{0/4}(\ell\theta) [P_{\rm E}(\ell)\pm P_{\rm B}(\ell)]\;, 
\end{align}
where $P_{\rm E}(\ell)$ and $P_{\rm B}(\ell)$ are the E and B-mode power spectra, respectively. 
Therefore, $\xi_\pm$ cannot be used for B-mode based diagnostics of shear systematics.

\subsection{COSEBIs}

COSEBIs\footnote{ We use logarithmic COSEBIs throughout this paper, which are the more efficient of the two families of COSEBIs for a cosmological analysis.} are designed to separate E/B-modes cleanly and completely over a finite angular range \citep{SEK10}. 
They achieve this in an efficient manner, where only a few COSEBI modes are required to capture essentially the full cosmological information contained within a dataset.
In this analysis we use the first 5 COSEBI modes in accordance with the results of \cite{Asgari12}. 
COSEBIs are defined as  integrals over the 2PCFs, 
\begin{align}
\label{eq:COSEBIsReal}
 E_n^{ij} &= \frac{1}{2} \int_{\theta_{\rm min}}^{\theta_{\rm max}}
 \d\theta\,\theta\: 
 [T_{+n}(\theta)\,\xi^{ij}_+(\theta) +
 T_{-n}(\theta)\,\xi^{ij}_-(\theta)]\;, \\ \nonumber
 B_n^{ij} &= \frac{1}{2} \int_{\theta_{\rm min}}^{\theta_{\rm
     max}}\d\theta\,\theta\: 
 [T_{+n}(\theta)\,\xi^{ij}_+(\theta) -
 T_{-n}(\theta)\,\xi^{ij}_-(\theta)]\;,
\end{align} 
where $T_{\pm n} (\theta)$ are complete sets of filter functions defined such that they cleanly separate E/B-modes and remove all ambiguous modes which cannot be uniquely defined as either E or B. 
These filters are zero outside of their support \citep[see Fig.\;1 in][]{asgari/etal:2019a}. We use a discrete form of \Eqt\eqref{eq:COSEBIsReal} to estimate COSEBIs from the measured values of 2PCFs (\Eqt\ref{eq:xipm_meaure}). 
The theoretical predictions for COSEBIs can also be calculated from the convergence power spectra,
\begin{align}
\label{eq:EnBnFourier}
E_n^{ij} &= \int_0^{\infty}
\frac{\d\ell\,\ell}{2\pi}P^{ij}_{\mathrm{E}}(\ell)\,W_n(\ell)\;,\\ \nonumber
B_n^{ij} &= \int_0^{\infty}
\frac{\d\ell\,\ell}{2\pi}P^{ij}_{\mathrm{B}}(\ell)\,W_n(\ell)\;,
\end{align} 
where $W_n(\ell)$ are the Hankel transforms of $T_{\pm n}(\theta)$,
\begin{align}
\label{eq:Wn}
W_n(\ell) & =  \int_{\theta_{\rm{min}}}^{\theta_{\rm{max}}}\d\theta\:
\theta\:T_{+n}(\theta) \rm{J}_0(\ell\theta)\;, \nonumber \\ 
& = \int_{\theta_{\rm{min}}}^{\theta_{\rm{max}}}\d\theta\:
\theta\:T_{-n} (\theta) \rm{J}_4(\ell\theta)\;.
\end{align} 
The $W_n(\ell)$ are oscillatory functions that peak near $2\pi/\theta_{\rm max}$ and reach zero at small and large values of $\ell$ \citep[see Fig.\;2 in][]{Asgari12}.

\begin{figure*}
   \begin{center}
     \begin{tabular}{c}
     \includegraphics[width=0.65\hsize]{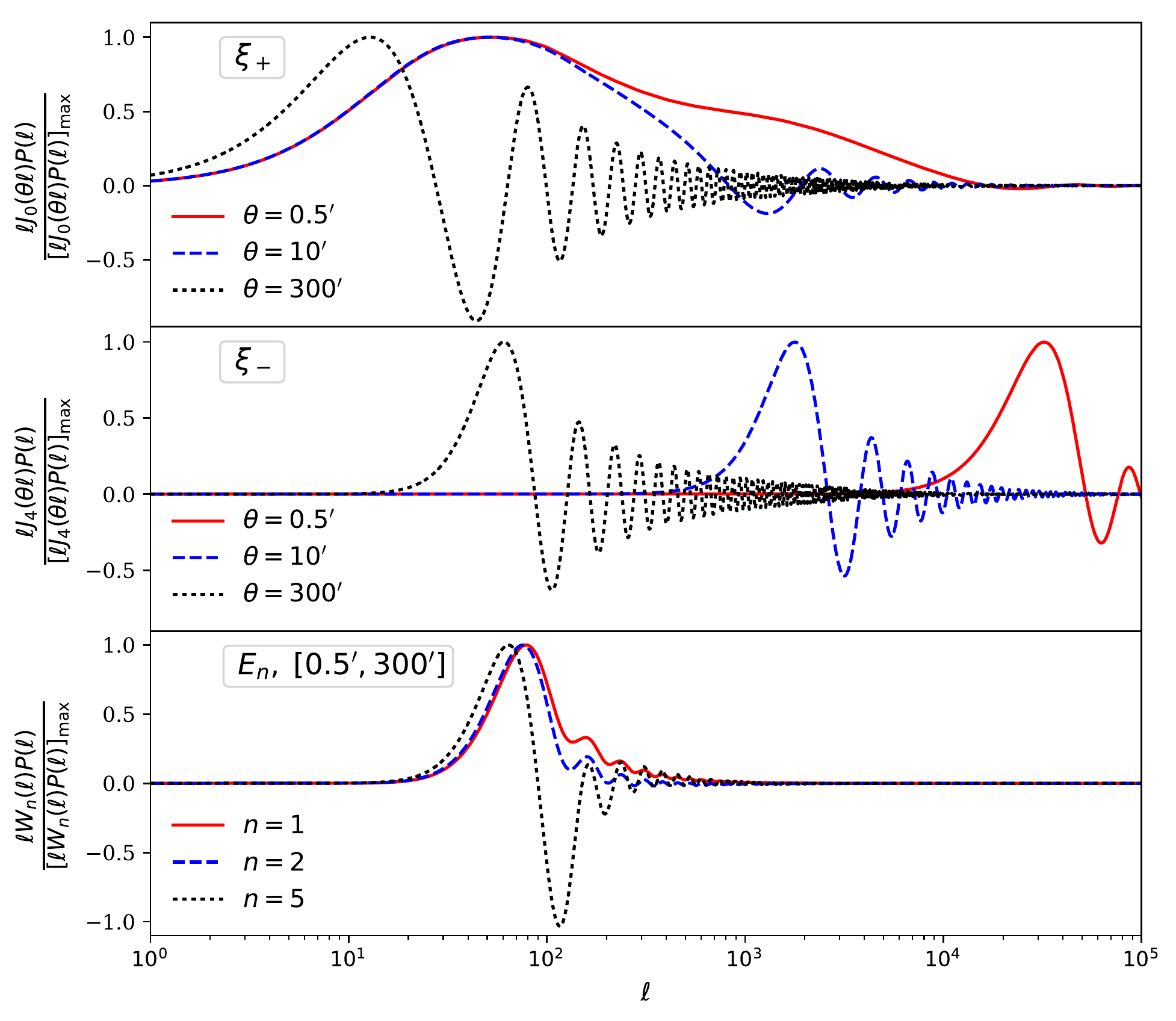}
      \includegraphics[width=0.35\hsize]{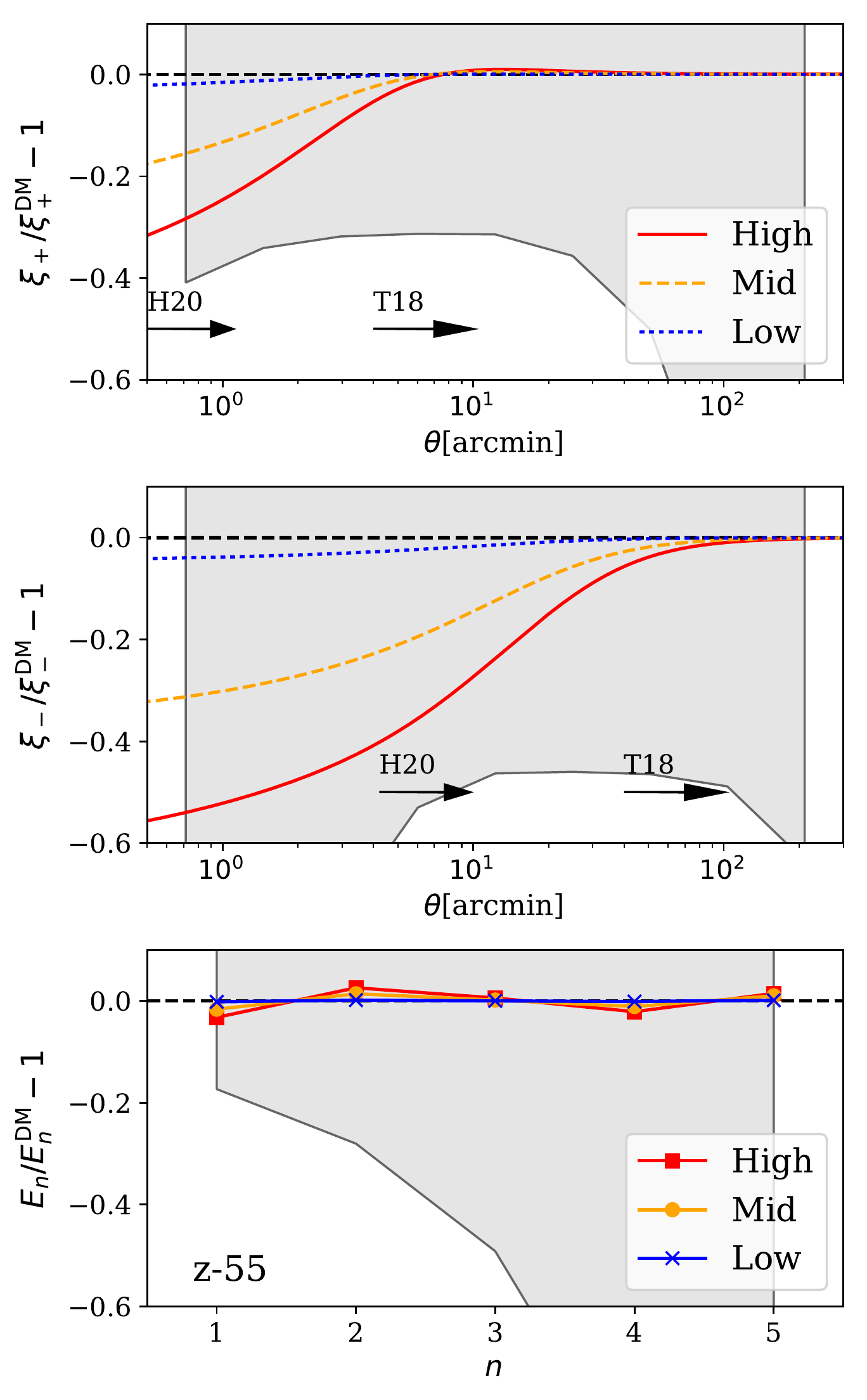} 
     \end{tabular}
   \end{center}
     \caption{\small{Comparison between $\xi_+$, $\xi_-$ and COSEBIs $E_n$. \textit{Left:} Integrands are shown for three angular distances, $\theta$, for $\xi_+$, $\xi_-$ and three modes for $E_n$ defined over the $\theta$-range $[0.5', 300']$. All integrands are normalised by their maximum value. \textit{Right:} Sensitivity of $\xi_\pm$ and $E_n$ to baryon feedback. Three baryon feedback cases from \hmcode are shown with different feedback parameters, High: $A_{\rm bar}=1$, Mid: $A_{\rm bar}=2$ and Low: $A_{\rm bar}=3$. All curves are normalised with respect to the dark matter only case with a baryon feedback of $A_{\rm bar}=3.13$. The grey shaded regions show the error bars for KV450. The $\xi_\pm$ errors are shown for the 9 logarithmic bins used in the KV450 analysis. The arrows show which scales were used in the primary KV450 (H20) and DES-Y1 analyses (T18).}}
     \label{fig:compare}
 \end{figure*}

\subsection{Comparison of 2PCFs and COSEBIs}
\label{sec:comparison}

In this section we explore the dependence of 2PCFs and COSEBIs on the convergence power spectrum at different multipoles, $\ell$. 
The left panels of \fig\ref{fig:compare} show the integrands in Eqs.\;\eqref{eq:xipm} and \eqref{eq:EnBnFourier} that translate the convergence power spectrum, $P_\kappa(\ell)$, into $\xi_\pm$ and COSEBIs, respectively. {Here and in the rest of this paper we assume that $P_{\rm B}(\ell)$ vanishes and set $P_\kappa(\ell)=P_{\rm E}(\ell)$.}
From top to bottom the panels denote $\xi_+$, $\xi_-$ and $E_n$. 
The integrands for $\xi_\pm$ are shown for three angular distances: $0.5'$ (solid red), $10'$ (dashed blue), and $300'$ (dotted black). 
For COSEBIs we show three modes: $n=1$ (solid red), $n=2$ (dashed blue), and $n=5$ (dotted black), all of which use the angular range of $[0.5', 300']$. 
This is the angular range employed for the KV450 primary cosmic shear analysis (H20), which also encompasses the range used by DES in T18. 
The amplitudes of the integrands are normalised with respect to their maximum values. 
The convergence power spectrum corresponds to the auto-correlation of the highest redshift bin used in the KV450 analysis while the cosmological parameters are those of its best fitting $\Lambda$CDM model. 
This figure shows that the region in $\ell$ over which the integrand is significantly non-zero is smaller for COSEBIs compared to either $\xi_+$ or $\xi_-$, which demonstrates that COSEBIs are more robust against systematics that affect either small or large $\ell$-scales. 

The reduced dependence of COSEBIs on high multipoles of the convergence power spectrum is illustrated in the right hand panels of \fig\ref{fig:compare}, which show the sensitivity of $\xi_+$ (top), $\xi_-$ (middle), and $E_n$ (bottom) to baryon feedback. 
The vertical axes show the fractional difference between a baryon feedback and a dark matter-only model for three baryon feedback cases. 
We use \hmcode\!\footnote{{The baryon feedback in \hmcode is calibrated against OverWhelmingly Large Simulations \citep[OWLS;][]{vandaalen/etal:2011}.}} \citep{mead/etal:2015} to model the feedback and vary its free parameter, $A_{\rm bar}$\footnote{For a single parameter baryon feedback model the 'halo bloating' parameter is coupled to $A_{\rm bar}$ via $\eta=0.98-0.12 A_{\rm bar}$.}, to produce a low feedback case with $A_{\rm bar}=3$ (blue), a medium feedback case with $A_{\rm bar}=2$ (orange), and a high feedback case with $A_{\rm bar}=1$ (red). 
Note that the dark matter-only case corresponds to $A_{\rm bar}=3.13$. 
The error on the measurements is shown as the grey region, assuming the autocorrelation of the highest redshift bin in KV450. 
The arrows in the upper panels show which angular ranges were used in H20 and T18. 

\fig\ref{fig:compare} shows how baryon feedback introduces significant suppression in the $\xi_+(\xi_-)$ signal by up to $30\%\:(40\%)$ on scales used in the KiDS analysis \citep[see also][]{semboloni/etal:2011}. 
Here we show the effect of baryon feedback on the autocorrelation of the highest tomographic bin for KV450. 
Galaxies in this bin are on average at a higher redshift compared to the lower tomographic bins and hence a given $\ell$-scale for the fifth bin probes larger physical scales compared to the other bins. 
As a result the signal in this tomographic bin is less affected by baryon feedback compared to the same angular scales measured for a lower redshift bin. In terms of signal-to-noise, however, the baryon feedback contribution to this bin is the most significant compared to all autocorrelated redshift bins.
In the lower right hand panel of \fig\ref{fig:compare} we show that COSEBIs signals for $[0.5', 300']$ are affected by baryon feedback by at most $3\%$. Choosing COSEBIs therefore mitigates against much of the uncertainty in the baryon feedback modelling, increasing the robustness of the analysis.

In H20 baryon feedback suppression was modelled using \hmcode with an informative prior on $A_{\rm bar}$ and a small scale cut of $\sim 4.2'$ for $\xi_-$. 
T18 took a more conservative approach of excluding most of the small-scale information from their analysis, selecting angular scales considered to be unaffected by baryon feedback by excluding $\theta$-scales that are affected by Baryon feedback by more than $2\%$ when comparing the dark matter only signal to the OWLS AGN simulations. The feedback intensity of these simulations lies somewhere in between the low and medium feedback cases shown. If we apply the T18 2\% effect criteria, we can conclude that the angular scale cuts advocated by T18 are no longer necessary to apply when using COSEBIs. 

\section{Data and analysis pipeline}

\label{sec:data}

\begin{table}
\centering
\caption{\small{Cosmological parameters and priors used in H20 \citep{hildebrandt/etal:2020} and T18 \citep{troxel/etal:2018a} setups. The first column shows the parameter and the second and third show the prior range adopted by KiDS and DES, respectively. Flat priors are indicated in square brackets, showing the edges of the prior and the Gaussian priors are shown as the mean $\pm$ the standard deviation on the parameter. The first 10 are cosmological parameters: $A_{\rm s}$ is the amplitude  (defined at $k_0=0.05\; {\rm Mpc}^{-1}$) and $n_{\rm s}$ is the spectral index of the primordial power spectrum. $\Omega_{\rm b}$, $\Om$, and $\Omega_{\rm CDM}$  are  the density parameters for baryonic, total and cold dark matter, respectively, while $\Omega_\nu$ is the neutrino density parameter. Finally $h$ is the dimensionless Hubble parameter. The next three parameters are the amplitude, $A_{\rm IA}$ and redshift evolution, $\eta_{\rm IA}$, parameters for the intrinsic alignment of galaxies and the baryon feedback parameter $A_{\rm bar}$ (called $B$ in H20). The final set of parameters capture the systematic effects, where $m_i$ is the multiplicative shear bias, $\sigma_m$ is the dispersion of $m_i$ and $\Delta z_i$ is the additive shifts in redshift distributions for each bin. In the H20 setup the effect of $m$ is applied to the data and its error is added to the covariance matrix (see \Eqt\ref{eq:sigmam}).
 $A_{\rm c}$ is the amplitude of the 2D additive bias explained in H20. { The model used to evolve the non-linear matter power spectrum is shown in the last row.} }}
\label{tab:setups}
\resizebox{\columnwidth}{!}{
\renewcommand{\arraystretch}{1.2}
\begin{tabular}{ c  c  c    }
         Parameter                            & H20                          & T18   \\ \hline\hline
$A_{\rm s}$                            &    $-$                            &  $[5\times 10^{-10}, 5\times 10^{-9}]$         \\ \hline
$\ln (10^{10} A_{\rm s})$       &    $[1.5, 5.0] $             &  $-$    \\ \hline
$n_{\rm s}$                            &    $ [ 0.7,1.3]        $             & $[0.87 ,    1.07]$ \\ \hline
$\Omega_{\rm b}$        &     $-$                                 & $[0.03 ,  0.07]$ \\ \hline
$\Omega_{\rm b} h^2$  &     $[0.019, 0.026]$           & $-$   \\ \hline
$\Om$                           &   $-$                                 &  $[0.1, 0.9]$       \\ \hline
$\Omega_{\rm CDM} h^2$  &   $ [0.01, 0.99 ] $            &  $-$         \\ \hline
$\Omega_{\nu} h^2$           & $-$                                        & $[0.0006, 0.01] $\\ \hline 
$\Omega_{\nu}$                 & $0.00131$                   &                      $-$     \\ \hline 
$h$                               &   $ [0.64, 0.82] $            &  $[0.55, 0.9]$       \\ \hline
$A_{\rm IA}$               &   $ [-6,6]       $                    & $[-5,5]$ \\ \hline
$\eta_{\rm IA}$        &    $0$                                        &  $[-5, 5]$ \\ \hline 
$A_{\rm bar}$              &    $[2, 3.13] $                &     None  \\  \hline \hline
$m_i$                         &     $-$                                    & $0.012 \pm 0.023$ \\ \hline
$\sigma_m$               &     $0.02$                                    & $-$ \\ \hline
$\Delta z_1$               &      $ 0.0\pm 0.039   $        & $0.001\pm 0.016$ \\ \hline
$\Delta z_2$              &        $0.0 \pm 0.023$     & $-0.019\pm 0.013$ \\ \hline
$\Delta z_3$              &        $0.0 \pm 0.026$    & $0.009\pm 0.011$ \\ \hline
$\Delta z_4$               &        $0.0 \pm 0.012$        & $0.018\pm 0.022$\\ \hline
$\Delta z_5$               &         $0.0 \pm 0.011$     & $-$    \\ \hline
$A_{\rm c}$                &      $1.01\pm 0.13 $              &  0  \\ \hline \hline
Non-linear model               &      \cite{mead/etal:2015}      & \cite{Takahashi12} \\ \hline
\end{tabular}
}
\end{table}

We measure COSEBIs for two sets of public data: KV450 and DES-Y1, and fit a flat $\Lambda$CDM model to these datasets. 
We are interested in constraints on $\Om$, the matter density parameter, $\sigma_8$, the standard deviation of present day linear matter perturbations in a sphere of radius 8 $h^{-1}$\,Mpc, 
and $S_8=\sigma_8 (\Om/0.3)^{0.5}$, a combination of parameters that the cosmic shear signal is most sensitive to.

We use {\sc CosmoSIS} \citep{cosmosis}, a modular cosmological parameter estimation code,
 modified to include modelling of the nonlinear power spectrum using the halo-model formalism of  \citet[\hmcode\!\!]{mead/etal:2015} and the COSEBIs pipeline (Eqs.\:\ref{eq:EnBnFourier}, \ref{eq:Wn} and \ref{eq:COSEBIsCov}).
Posteriors are derived using the {\sc emcee} sampler \citep{emcee}\footnote{emcee.readthedocs.io} for KV450  and the {\sc MultiNest} sampler \citep{multinest} for DES-Y1 in accordance with their original analysis.

When comparing the calculation of COSEBIs from \Eqt\eqref{eq:COSEBIsReal} and \Eqt\eqref{eq:EnBnFourier} we find agreements of better than $0.03\%$ \citep[see figure A.1 in][]{asgari/etal:2017}. To achieve this accuracy we first calculate the  $W_n(\ell)$ functions for a large range of $\ell$ with high resolution and save them to file for further use (1 million logarithmically binned samples between $\ell=1$ and $\ell=10^5$ for the angular range of $0.5'-300'$.). Therefore, for each angular range we need to calculate the $W_n(\ell)$ only once. After this stage we can load the weight functions to memory for calculating COSEBIs which is of similar speed to 2PCFs.

The linear power spectrum is calculated with {\sc camb} \citep{camb2000,camb12} and either \hmcode (H20 setup) or \cite{Takahashi12} (T18 setup) are used to predict the non-linear evolution of the matter power spectrum. 
A Limber approximation according to \Eqt\eqref{eq:limber} is then applied to the matter power spectrum to obtain the convergence power spectrum.  
The effect of the intrinsic alignment of galaxies is added to the predictions using the \cite{Bridle07}\footnote{{bk{\_}corrected}} model. 
The parameters and priors used in our analysis are shown in \tab\ref{tab:setups} for two setups: H20 and T18 referring to
the setups used in \cite{hildebrandt/etal:2020} and \cite{troxel/etal:2018a}, respectively. 
We use H20 as our fiducial setup in this analysis\footnote{With the exception of the lower bound on $\ln(10^{10} A_{\rm s})$.  H20 used 1.7 instead of 1.5 that we use here.} . This setup results in more conservative results as shown in \cite{joudaki/etal:2019} and also includes \hmcode which provides a better match to simulations and includes the effect of baryon feedback.

COSEBIs are estimated by using trapezoidal integrations over the measured 2PCFs (see \Eqt\ref{eq:COSEBIsReal}). 
The 2PCFs are measured using {\sc athena} \citep{KilbingerAthena14} with $1,200,000$  linear $\theta$-bins in the range of $[0.5', 300']$. 
The number of $\theta$-bins needed to get an unbiased estimate for COSEBIs has been investigated in \cite{asgari/etal:2017}.
The covariance of COSEBIs is calculated analytically for the best fitting parameters of each dataset. The shape noise contribution to the covariance matrix is estimated using the number of galaxy pairs and the effective ellipticity dispersion estimated from the data. See \App\ref{app:covariance} for further details.

\subsection{KV450}
\label{sec:KV450}

The Kilo-Degree Survey (KiDS) (\citealt{Kuijken15}, \citealt{dejong/etal:2017}) combined with VIKING \citep{edge/etal:2013} allow for 9-band, $ugriZYJHK_s$, photometric observations of galaxies that can be used for cosmic shear analyses (H20, \citealt{ wright/etal:2019}). 
These photometric bands cover a wide range of frequencies which improves the accuracy of the redshift estimates and decrease the fraction of catastrophic outliers compared to the more limited set of photometric bands used in \cite{hildebrandt/etal:2017}. 
The galaxies are first divided into five photometric redshift bins using {\sc bpz} \citep{bpz2000,bpz2004}: $z_{\rm phot}\in(0.1, 0.3]$, $z_{\rm phot}\in(0.3, 0.5]$, $z_{\rm phot}\in(0.5, 0.7]$, $z_{\rm phot}\in(0.7, 0.9]$, and $z_{\rm phot}\in(0.9, 1.2]$. 
Then the redshift distribution of galaxies in each bin is calibrated against spectroscopic surveys.
This spectroscopic sample crucially includes deep surveys which are necessary to calibrate the faint galaxies at high redshifts, which contribute strongly to the final signal. We follow the calibration method used in \citet["DIR"]{hildebrandt/etal:2017}. 
The DIR method uses a $k$-nearest neighbour algorithm to assign weights to galaxies in the spectroscopic sample, matching their weighted distribution to the photometric sample used in the cosmic shear analysis \citep{lima/etal:2008}.
This method can recover the true distribution of the galaxies as long as the full photometric sample is represented in the spectroscopic sample \citep{wright/etal:2019b}.

Here we analyse the KV450 data, processed by {\sc theli} \citep{Erben13} and Astro-WISE \citep{begeman/etal:2013}. 
Galaxy ellipticities are measured in the $r$-band with \emph{lens}fit \citep{Miller13} and have been calibrated with image simulations \citep{kannawadi/etal:2019}, such that the multiplicative shear biases are consistent with zero within their estimated dispersion of $\sigma_m=0.02$. 
COSEBIs are estimated from the angular range of $[0.5', 300']$ for KV450.
The effective area associated with this dataset is equal to $341.29$ deg$^2$, which we use to scale our covariance matrix. 
We use the same free parameters and priors used in the KV450 cosmic shear analysis of H20, with the exception of the `constant $c$-term offset' parameter, since COSEBIs are unaffected by this $c$-term, thus reducing the total number of varied parameters to 13 (see \tab\ref{tab:setups}).

\subsection{DES-Y1}
\label{sec:DESy1}

We use the first year of public Dark Energy Survey data \citep[DES-Y1,][]{DES_DataRelease1}\footnote{https://des.ncsa.illinois.edu/releases/y1a1} with additional masks applied to match the cosmic shear analysis of T18. 
This results in an effective area of 1321 deg$^2$, with galaxies observed in four photometric bands $griz$ \citep[see][]{morganson/etal:2018,flaugher/etal:2015}. 
The ellipticties of galaxies are measured using two methods: \im  and \mcal \citep{zuntz/etal:2014,sheldon_huff:2017,Zuntz/etal:2018}, with \mcal resulting in a larger galaxy sample which we use for our analysis. 

DES galaxies are binned into 4 redshift bins: $z_{\rm phot}\in(0.2 , 0.43)$,  $z_{\rm phot}\in(0.43 ,0.63)$,  $z_{\rm phot}\in(0.63 , 0.9)$,  $z_{\rm phot}\in(0.9 , 1.3)$. 
The fiducial galaxy redshift distributions are measured using {\sc bpz} \citep{bpz2000} and calibrated with "COSMOS-2015" \citep{COSMOS2015}.   This calibration results in a shift to the mean of the {\sc bpz} redshift distributions through the prior on $\Delta z_i$ as shown in \tab\ref{tab:setups} \citep[see][for details]{Hoyle/etal:2018}. 

For the DES data, we measure COSEBIs using the angular range of $[0.5', 250']$ which is the full angular range shown in T18. 
In \cite{asgari_heymans:2019}, it was shown that the B-modes of the \mcal catalogue are consistent with zero for $[0.5', 250']$, demonstrating that there is no evidence for B-mode producing systematics (e.g. PSF residuals) over this range. 
Additionally, in \sect\ref{sec:comparison} we compared $\xi_\pm$ and COSEBIs over a similar angular range and showed that COSEBIs are less sensitive to small physical scales and hence the effects of baryon feedback.
Consequently, there is no motivation to adopt the conservative approach of T18 and remove the small angular scales in our analysis.

We explore three setups for the DES-Y1 analysis:
\begin{itemize}
\item T18 setup: We use the DES {\sc bpz} redshift distributions and the parameters and priors used in T18, resulting in 16 free parameters, including four multiplicative bias parameters, as listed in \tab\ref{tab:setups}. 
For this setup we do not include the $\sigma_m$ error in the covariance (see \Eqt\ref{eq:sigmam}). \\

\item H20 setup with DES {\sc bpz} redshift distributions: Here we are interested in quantifying the effect of parameter selection and priors on the analysis. We therefore use the H20 setup for this analysis, keeping the redshift distributions and priors unchanged. 
This results in 11 free parameters, 2 less than the KV450 case  (\sect\ref{sec:KV450}), since DES-Y1 has one fewer redshift bin (we use the DES priors on $\Delta z_i$) and we do not include the 2D $c$-term (see \tab\ref{tab:setups}). In this case we apply the error on the multiplicative bias using \Eqt\eqref{eq:sigmam} and $\sigma_m=0.023$. The mean $m_i=0.012$ correction is also taken into account in the theoretical prediction. \\

\item H20 setup with DIR calibrated redshift distributions: Using 9-band KiDS data, H20 found that using COSMOS-2015 to calibrate photometric redshifts resulted in an underestimation of the redshift distributions of the  galaxies, in comparison to the distributions directly calibrated using 9-band spectroscopic data. \citet{joudaki/etal:2019} apply the same direct spectroscopic calibration technique (`DIR')  to the 4-band DES-Y1 data (including the selection responses from the \mcal method), utilising overlapping spectroscopic data from zCOSMOS, VVDS (VIMOS VLT Deep Survey) 2h and 22h, DEEP2-2h and CDFS (Chandra Deep Field South)  (see H20 and the discussion in \citealt{joudaki/etal:2019} for details). 
The DIR redshift distributions for DES-Y1 are shown in \cite{joudaki/etal:2019}. 
The means of the redshift distributions with the 4-band DIR calibration are higher than the COSMOS-2015 calibrated {\sc bpz} distributions.
The parameters and priors match the `H20 setup with {\sc bpz}' case and the only change here is the redshift distributions and their priors: $\Delta z_1=0 \pm 0.008$, $\Delta z_1=0 \pm 0.014$, $\Delta z_1=0 \pm 0.011$ and $\Delta z_1=0 \pm 0.009$. The uncertainties on the $\Delta z_i$ are determined from a spatial bootstrap resampling of the calibration sample.

\end{itemize}

\begin{figure*}
   \begin{center}
     \begin{tabular}{c}
      \includegraphics[width=\hsize]{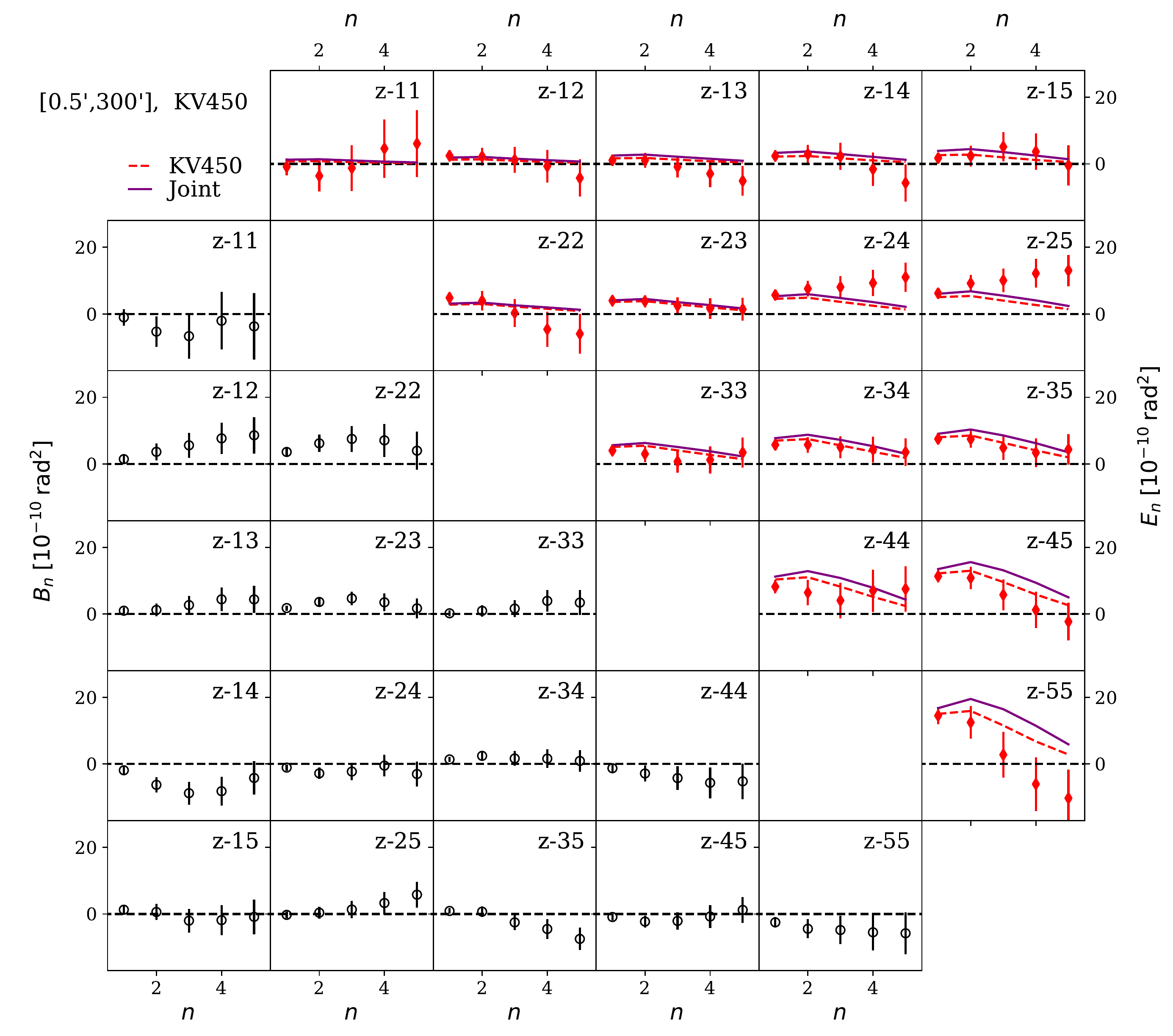}
     \end{tabular}
   \end{center}
     \caption{\small{COSEBIs measurements and their expected values for KV450. 
     The E-modes are shown in the upper triangle and the B-modes in the lower triangle. 
     Each panel corresponds to a redshift bin pair as indicated in its corner.    
     Theoretically expected values are shown for the best fitting parameters for KV450 (red dashed) and its combination with DES-Y1 (solid purple).  
     The COSEBI modes are discrete and the points are connected to each other purely for visual purposes. 
     The error on the B-modes is calculated assuming that the only contribution is shape noise, while for the E-modes we take all Gaussian terms as well as the super sample covariance into account (see \App\ref{app:covariance}). Neighbouring COSEBI modes are correlated and the goodness-of-fit of the model cannot be established by eye (see \tab\ref{tab:bestfit}).}}
     \label{fig:COSEBIsKV450}
 \end{figure*}

\begin{figure*}
   \begin{center}
     \begin{tabular}{c}
      \includegraphics[width=\hsize]{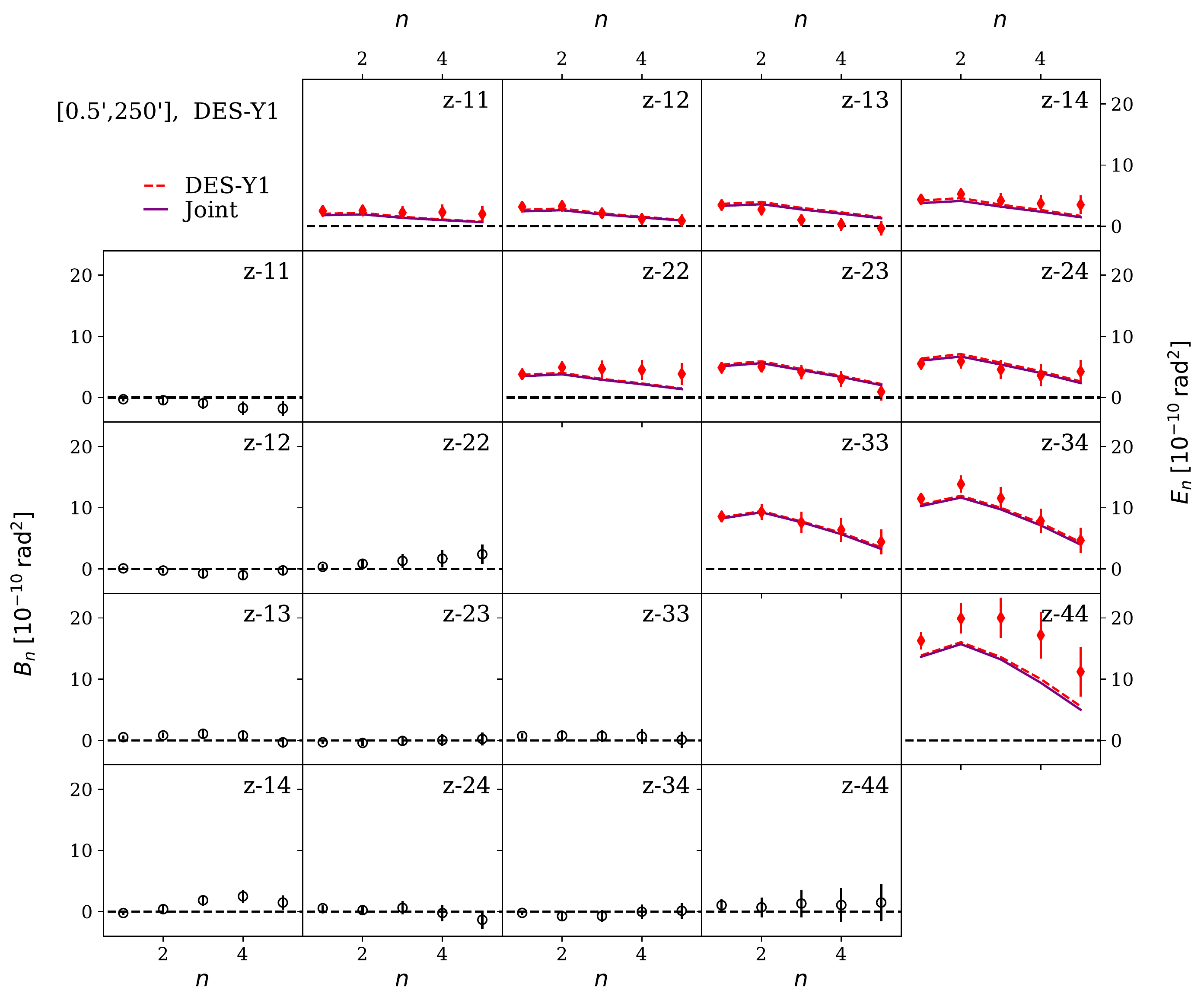}
     \end{tabular}
   \end{center}
     \caption{\small{COSEBIs measurements and their expected values for DES-Y1. 
     Similar to \fig\ref{fig:COSEBIsKV450}, E and B-modes are shown on the upper and lower triangles, respectively. 
     The redshift bin is indicated in the corner of each panel. 
    Theoretically expected values for the best fitting parameters using DES-Y1 (red dashed) and the joint analysis with KV450 (solid purple) are shown as curves, although COSEBI modes are discrete. Both theoretical models fit the data well (see \tab\ref{tab:bestfit}). As the data points are correlated the goodness-of-fit cannot be inspected visually. }}
     \label{fig:COSEBIsDESy1}
 \end{figure*} 

\subsection{Joint Analysis}
\label{sec:joint}

Beside the individual analyses of DES-Y1 and KV450, we also conduct a joint analysis assuming no correlations between the two datasets,
since there is no overlap between the DES-Y1 and KV450 data on the sky.

For the joint analysis, we employ the H20 setup and use DIR calibrated redshifts for both datasets (see \tab\ref{tab:setups}).
All cosmological and astrophysical parameters are shared between the two datasets, but their systematic parameters are allowed to vary individually. 
These systematic parameters are the additive shifts for the redshift distributions (five parameters for KV450 and four for DES-Y1), as well as the 2D $c$-term amplitude for KV450. Note that the uncertainty on the multiplicative bias  parameters are absorbed into the covariance matrix for each survey (see \Eqt\ref{eq:sigmam}). As we will see in \sect\ref{sec:results} when using this setup there is excellent agreement between the individual constraints of KV450 and DES-Y1, allowing us to carry out the joint analysis.
{ \cite{joudaki/etal:2019} also followed a similar setup for their joint analysis, but kept the intrinsic alignment parameters separate. However, they show that this choice does not affect the cosmological parameters significantly. }

\section{Results}
\label{sec:results}

\begin{figure}
   \begin{center}
     \begin{tabular}{c}
      \includegraphics[width=\hsize]{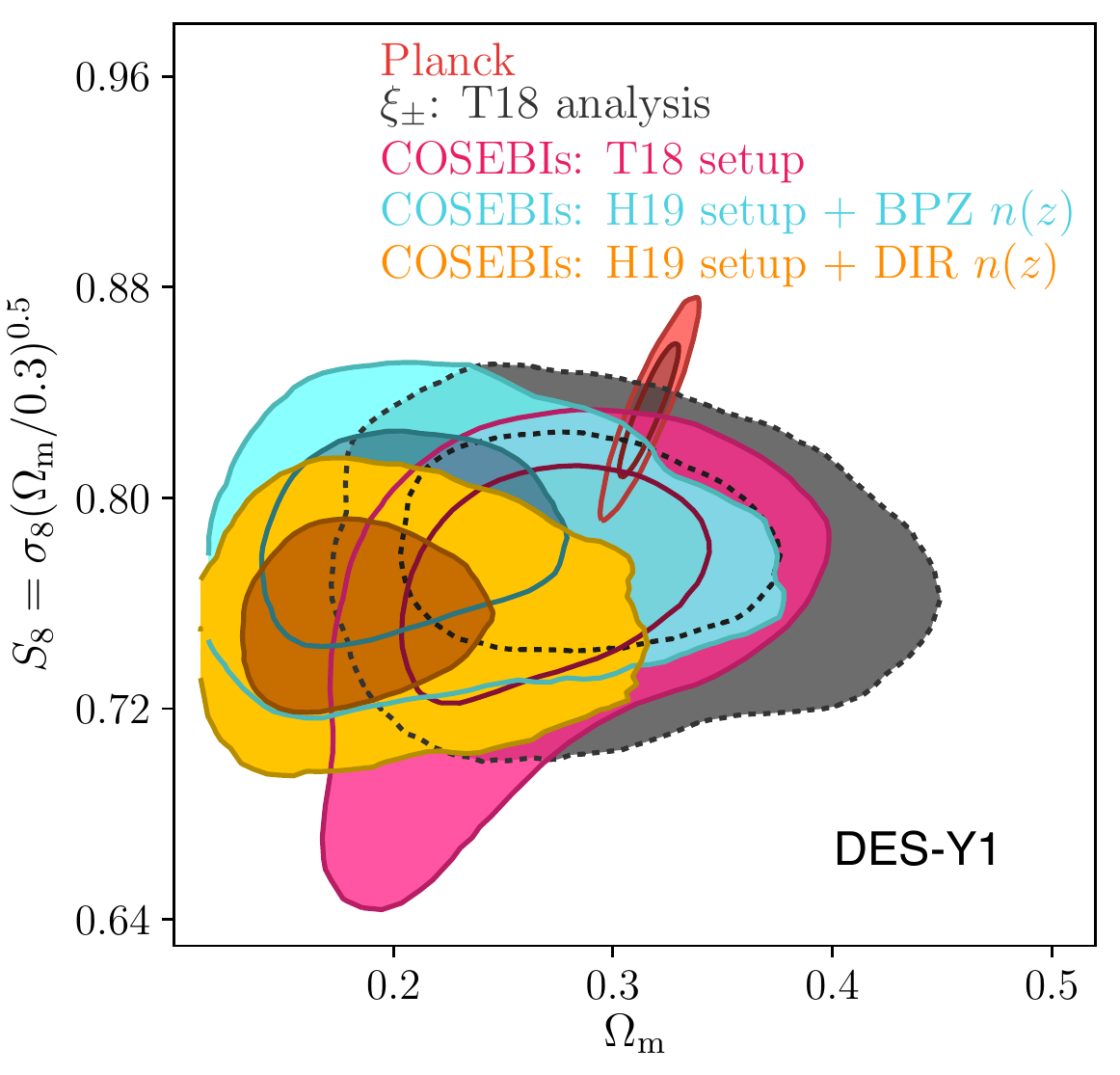}
     \end{tabular}
   \end{center}
     \caption{\small{Constraints on $S_8$ and $\Om$.  DES-Y1 analysis results with COSEBIs using the T18 setup (magenta), the H20 setup with {\sc bpz} redshift distributions (cyan) and the H20 setup with DIR calibrated redshift distributions (orange). 
     The grey contours belong to the fiducial analysis in \cite{troxel/etal:2018a} and the red contours show Planck Legacy results (TT,TE,EE+lowE).}}
     \label{fig:S8Om}
 \end{figure}

 \begin{figure*}
   \begin{center}
     \begin{tabular}{c}
      \includegraphics[width=0.5\hsize]{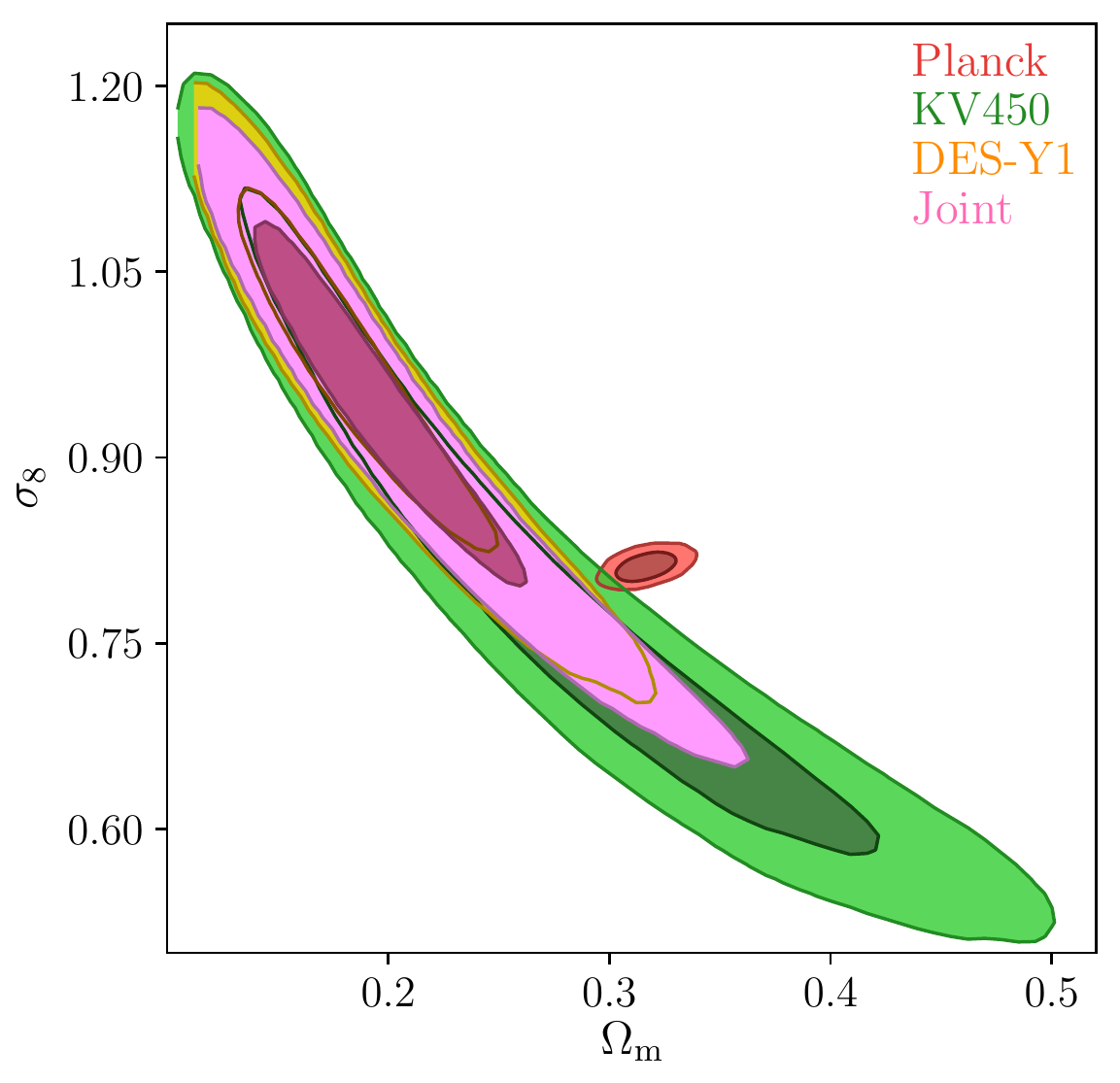}
      \includegraphics[width=0.5\hsize]{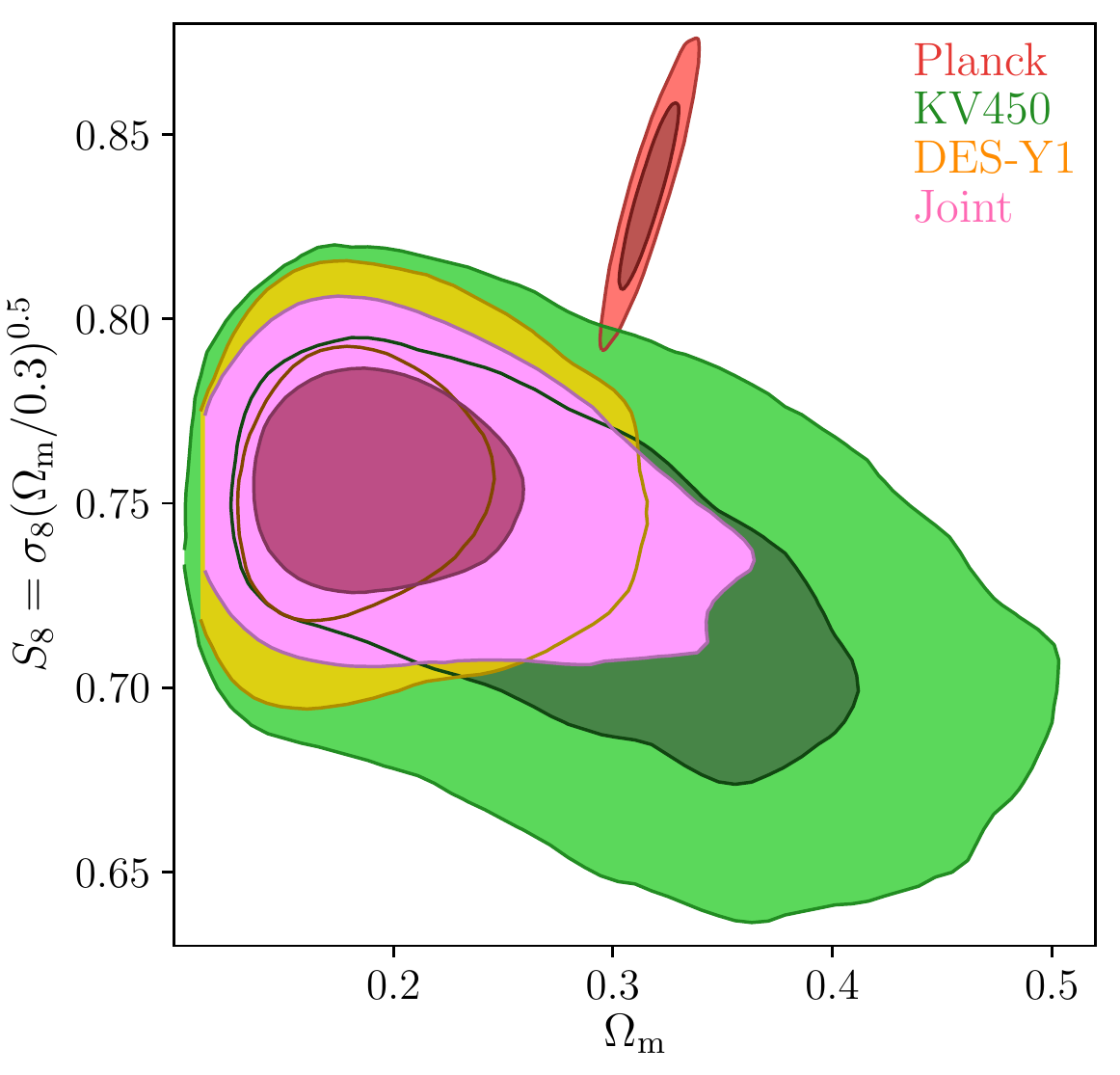}
     \end{tabular}
   \end{center}
     \caption{\small{KV450 analysis results with COSEBIs (green), DES-Y1 results with H20 setup and DIR spectroscopically calibrated redshifts (orange-yellow) and their joint analysis (pink).   Red contours show the Planck legacy results for TT,TE,EE+lowE. Constraints on $\sigma_8$ and $\Om$ are shown in the left panel, while the right panel shows results for $S_8=\sigma_8(\Om/0.3)^{0.5}$ and $\Om$. KV450 constraints for $S_8=0.737^{+0.036}_{-0.038}$, DES-Y1 $S_8=0.755\pm{0.023}$ and their joint constraint is $S_8=0.755^{+0.019}_{-0.021}$ which is in $3.2\sigma$ tension with the Planck constraints $S_8=0.834\pm 0.016$.}}
     \label{fig:sig8Om}
 \end{figure*} 

\begin{table}
\centering
\caption{\small{$\chi^2$ (second column) and $p$-values (fourth column) for the best fitting parameters, given the degrees-of-freedom, DoF= number of data points $-$ number of free parameters (third column). 
The first column shows which dataset is used, with joint corresponding to the combination of DES-Y1 and KV450.  
In the last column we show the marginal maximum posterior value of $S_8=\sigma_8(\Om/0.3)^{0.5}$ for each case. 
Note that the effective degrees-of-freedom is larger than the value shown here due to correlations between the parameters and the existence of informative priors.  All results correspond to analyses of COSEBIs E-modes with the H20 setup and DIR redshifts.}}
\label{tab:bestfit}
\renewcommand{\arraystretch}{1.5}
\begin{tabular}{ c  c  c  c  c }

             & $\chi^2$ & DoF            & $p$-value       &   $S_8\pm 1\sigma$        \\ \hline
DES-Y1  &   57.8    &  $50-11$       &  0.03              &  $0.755\pm{0.023}$         \\  \hline
KV450 &    67.8      &  $75-13$       &  0.28             &   $0.737^{+0.036}_{-0.038}$      \\  \hline
Joint     &   130       &  $125-17$     &  0.08            &   $0.755^{+0.019}_{-0.021}$  
\end{tabular}
\end{table}

Here we show results for our analysis of the KV450 and DES-Y1 data introduced in \sect\ref{sec:data} with COSEBIs explained in \sect\ref{sec:method}. 
Figs.\:\ref{fig:COSEBIsKV450} and \ref{fig:COSEBIsDESy1} present the measured COSEBIs for KV450 ($\theta\in[0.5', 300']$) and DES-Y1 ($\theta\in[0.5', 250']$). 
The E-modes are shown in the upper triangle while the B-modes can be seen in the lower triangle. 
The first five COSEBI modes for each redshift bin pair are used in the cosmological analysis and are also shown here. 
The COSEBI modes are discrete and the first few modes are sufficient to capture the full cosmological information \citep{Asgari12}\footnote{Including the first 10 modes in the current analysis yielded no significant improvements to the marginalised constraints on $\sigma_8$ and $\Om$.}. 
The red dashed curves show the predictions for COSEBIs using the best fitting parameters deduced from the analysis of each survey separately, while the purple solid curves show the same but using the best fitting parameters from the joint analysis. 
In both cases the analysis is performed using the H20 setup (see \tab\ref{tab:setups}) and DIR spectroscopically calibrated redshift distributions. 

Using the first few B-modes may not be sufficient to assess their significance for a survey, as some systematics affect E/B-modes at different scales \citep{asgari/etal:2019a}. 
The B-mode analysis of the DES-Y1 data was presented in \cite{asgari_heymans:2019}, while the B-modes in KV450 were analysed in H20, both showing insignificant levels of B-modes, which allows us to safely use the data.

Neighbouring COSEBI modes are strongly correlated, making it difficult to assess the goodness-of-fit of the model by eye. 
In the case of B-modes, there are no fitting parameters, such that we can use the $\chi^2$ value with the number of data points as the degrees-of-freedom to calculate a $p$-value which shows the probability to exceed the given $\chi^2$ value and evaluate the significance of the B-modes. 
This however becomes more complicated for the E-modes, where correlated parameters with informative priors are fitted to the data \citep[see, for example,][]{handley_lemos:2019}. 
If we assume that the degrees-of-freedom of the analysis is equal to the number of data points minus the number of parameters, we arrive at the minimum value for the degrees-of-freedom. 
This minimum value is valid for the case where no informative priors are applied and all parameters are independent. 
If the predictions provide a good fit to the data assuming this minimum value for the degrees-of-freedom then it will also be a good fit for the true number of degrees-of-freedom, which is larger in general.
With this assumption in mind we quantify the goodness-of-fit of the best fitting models to each dataset with $p$-values for the best fitting parameters given in \tab\ref{tab:bestfit}. 
The first column shows which dataset is analysed, the second, third and fourth columns show the values of the $\chi^2$, the degrees-of-freedom = number of data points $-$ number of parameters, and their corresponding $p$-values, respectively. 
The final column shows the marginal maximum posterior value for $S_8$ including its $1\sigma$ errors. 
We see that all $p$-values are above $0.01\approx 2.3\sigma$, showing that the model is a good fit to the data in all cases {($p=0.03$ corresponds to $1.88\sigma$).}
Note that these are lower bounds on the $p$-values and we expect them to be larger than this since the effective number of parameters in our analysis is considerably smaller than the number of free parameters. 
Comparing the $S_8$ values in the last column, we see that they are in good agreement within their associated errors.  The internal consistency of the data can be tested in a Bayesian framework using the methods in \cite{kohlinger/etal:2019}, applied to KV450 in H20 showing no inconsistencies in the data.

\fig\ref{fig:S8Om} shows the DES-Y1 contour plots for $S_8$ and $\Om$ with the Planck Legacy results shown in red \citep[TT,TE,EE+lowE,][]{Planck2018}\footnote{The chains are plotted with {\sc ChainConsumer} \cite{Hinton2016}: samreay.github.io/ChainConsumer}. 
Here we show the analysis of DES-Y1 data using the three setups introduced in \sect\ref{sec:DESy1}. 
The grey dashed contours show the result from the cosmic shear analysis of T18, which used $\xi_\pm$ with different scale-cuts for each pair of redshift bins, to primarily avoid the effects of baryon feedback. 
The magenta contours show our analysis of DES-Y1 data with the same setup as T18, but using COSEBIs over the full angular range of $[0.5', 250']$. 
The cyan contours show the effect of moving from the T18 setup to H20, retaining the DES {\sc bpz} redshift distribution. Finally the orange contours show the effect of using the DIR calibrated redshift distributions with the H20 setup. 

Since we use a larger angular range, we obtain more information from the DES-Y1 data compared to T18. 
For the T18 setup, the COSEBIs confidence regions in the $S_8-\Om$ plane are $20\%$ smaller than the T18 analysis, 
with $S_8 =0.779^{+0.018}_{-0.038}$.
The long tail towards smaller $S_8$ values (\fig\ref{fig:COSEBIsDESy1}) is introduced by a tendency towards small values of the intrinsic alignment parameter, $A_{\rm IA}$. 
There are indications for this tendency in the correlation function analysis as well. 
Moving to the H20 setup results in an increased $S_8$ value of $0.784^{+0.025}_{-0.024}$, which is primarily due to the change in the modelling of the non-linear matter power spectrum (\citealt{Takahashi12} used in T18 and \citealt{mead/etal:2015} with baryon feedback used in H20) and fixing $\Omega_\nu$.
We also see that going from the {\sc bpz} redshift distributions to DIR redshift distributions results in a $\sim1.2\sigma$ shift to smaller values, increasing the difference between Planck and DES-Y1 in agreement with the analysis of \cite{joudaki/etal:2019}. 
We find $S_8= 0.755\pm 0.023$ for the orange contours which is $17\%$ tighter than the T18 constraints.

This increased statistical power may at first seem contradictory. We argue that COSEBIs are insensitive to small physical scales and we may therefore wrongly conclude that the addition of small angular scales would not affect the COSEBIs measurement. In \fig\ref{fig:compare} we see that the 2PCFs for small angular scales still carry information from Fourier scales that COSEBIs are sensitive to. Including these $\theta$-scales, therefore, impact the COSEBIs signal. In addition the inclusion of the smaller $\theta$-scales increases the number of galaxy pairs, decreasing the shot-noise for the measured COSEBIs signal.

\fig\ref{fig:sig8Om} shows the COSEBIs contours for the KV450 analysis in green and the DES-Y1 analysis in orange (same as the orange contours in \fig\ref{fig:S8Om}). 
We also show the results of the joint analysis of these datasets in pink (see \sect\ref{sec:joint}).  
All three analyses have adopted the H20 setup and DIR calibrated redshift distributions. The left panel shows the constraints in the $\sigma_8$-$\Om$ plane, while the right panel presents the results for $S_8$ and $\Om$.
The differences between the KV450 contours shown in H20 and the green contours here are negligible, $3\%$ tighter constraints with COSEBIs with the same best fit value for $S_8$ 
(see \fig\ref{fig:KV450_banana}).

In \fig\ref{fig:sig8Om} we can see that the $2\sigma$ contours of the joint COSEBIs analysis and Planck chains do not touch. The joint cosmic shear constraint on $S_8$ is $0.755^{+0.019}_{-0.021}$ which is in $3.2\sigma$ tension with Planck given a flat $\Lambda$CDM model. 
We remind the reader that DES-Y1 contours here are tighter than the T18 constraints, since we use a wider angular range in our COSEBIs analysis (see \tab\ref{tab:bestfit} for the constraints on $S_8$). 

We have so far only quantified the tension between our cosmic shear analysis and Planck for $S_8$. 
In \App\ref{app:triangle} we show constraints on the sampled cosmological and astrophysical parameters for our fiducial cosmic shear analysis and conclude that there is no tension with other cosmological results when we consider the marginal posterior distributions for the rest of the parameters. 
Differences between the results of cosmological analysis can also be quantified in the full parameter space sometimes resulting in a more significant tension \citep[]{charnock/etal:2017, raveri_hu:2019}. Since the constraints on $\Om$and $\sigma_8$, in particular, are prior dependent we do not attempt to quantify the tension in the full parameter plane. 

\subsection{Impact of baryons}

In various parts of this paper we have mentioned the reduced impact of baryon feedback on the COSEBIs signal in comparison to the 2PCFs. Here we summarise and elaborate upon the key points.

In \fig\ref{fig:compare} we showed that for a high baryon feedback case the COSEBIs signals are affected by only $3\%$, while the signal for $\xi_+(\xi_-)$ are reduced by up to $30\% (55\%)$ over the same angular range. The high baryon feedback case that we considered ($A_{\rm bar}=1$) has been excluded by observations \citep[see for example][and references therein]{chisari/etal:2018}. Currently the largest realistic feedback is believed to match the OWLS hydrodynamical simulations. The closest feedback parameter that corresponds to this simulation is $2.1\lesssim A_{\rm bar} \lesssim 2.5$. As a result H20 set the minimum value of their  baryon feedback parameter to $A_{\rm bar}=2$ which corresponds to the medium feedback case in \fig\ref{fig:compare}. 

T18 used the sensitivity of the signal to set a criteria for the $\theta$-scales to exclude. A more informative criteria is to look at the impact of baryon feedback on the 
signal-to-noise $\Delta\nu\equiv$ |{Signal({\small DM-only})-Signal({\small with Feedback})}|/{Error}. 
In terms of signal-to-noise for the medium feedback case, COSEBIs are affected by a maximum of $\Delta\nu=12\%$, while $\xi_\pm$ are affected by up to $\Delta\nu=46\%$ on the angular range of $[0.5',300']$. When using the angular cuts employed in T18, the maximum impact of baryon feedback decreases for $\xi_\pm$ to $\Delta\nu=18\%$. 
T18 demonstrated that even with their conservative angular cuts on the data, the inclusion of baryon feedback according to OWLS AGN simulations shifts their value of $S_8$ higher by $0.6\sigma$ compared to their fiducial dark matter only case. Repeating their fixed OWLS-like analysis with COSEBIs we find a similar result of a $0.7\sigma$ shift in $S_8$ values. Our fiducial case where we allow the baryon feedback parameter to vary, increases $S_8$ by $0.4\sigma$ compared to the dark matter only case. 
In conclusion, we have shown that using COSEBIs allows us to include smaller angular scales, while retaining the same level of insensitivity to baryon feedback as was imposed by T18 on their 2PCFs. Nevertheless, we marginalise over the baryon feedback parameter in our fiducial analysis, since even though its impact on the signal is small, it has an important effect on the constrained parameters.   

We do not see the same information gain comparing our COSEBIs analysis to the H20 results.
The angular range used in H20, considering the combination of $\xi_\pm$, is the same as the angular range we use here with COSEBIs. In the absence of baryon feedback, if both $\xi_\pm$ are defined on the same angular range as COSEBIs, their combination will have more cosmological information. Using \hmcode to marginalise over the effect of baryon feedback, however, results in effectively excluding all small physical scale information (large $\ell$-scales in \fig\ref{fig:compare}) that would have otherwise provided additional information from $\xi_\pm$.

\section{Conclusions}

\label{sec:conclusions}

In this paper we presented cosmic shear constraints on $S_8=\sigma_8(\Om/0.3)^{0.5}$, using COSEBIs measurements on KV450 and DES-Y1 data. 
We analysed each dataset separately and found good agreement with their primary analyses that employed shear two-point correlation functions. We homogenised the priors, nonlinear modelling and redshift calibration between these two datasets and finally combined them  assuming no cross-correlations to get joint constraints on the parameters. 
Both H20 and T18 made scale cuts in their analysis to avoid small scale effects, such as baryon feedback. 
T18 included conservative cuts, removing the necessity to model and marginalise over baryon feedback. 
Here we showed that COSEBIs are considerably less sensitive to small physical scales and are therefore less affected by the impact of baryon feedback on the matter power spectrum, in contrast to $\xi_\pm$. 
In addition, COSEBIs, much like $\xi_-$, are insensitive to a constant shear bias. 
Consequently, we were able to use COSEBIs measurements on the angular range of $[0.5', 250']$ and $[0.5', 300']$ for DES and KiDS, respectively.
This extended range of scales tightened the confidence region for $S_8$ derived from DES-Y1 data by $ 17\%$ compared to the analysis of T18. 
For KV450 we use the same angular range as H20 {when considering the combination of both 2PCFs}, $[0.5', 300']$, and therefore find very similar results with a reduction of $3\%$ for $S_8$ confidence regions with our COSEBIs analysis.

For the DES-Y1 analysis we explored three setups: 1) T18 setup with DES {\sc bpz} redshift distributions, 2) H20 setup with DES {\sc bpz} redshift distributions and 3) H20 setup with DIR spectroscopically calibrated redshift distributions. 
The first case corresponds to the setup used in \cite{troxel/etal:2018a} and the H20 setup matches \cite{hildebrandt/etal:2020}. 
H20 demonstrated that if the redshift distributions of galaxies are not calibrated with sufficiently deep spectroscopic surveys, the resulting cosmological analysis {can be shifted towards larger values of $S_8$}. This can be true even if we allow for uncorrelated additive redshift calibration parameters for each tomographic bin in the analysis, as the calibration error is typically correlated between redshift bins.
We therefore chose to analyse the data with the DIR spectroscopically calibrated redshift distributions. 
For our joint analysis, we needed to make a decision for the set of parameters and priors to be sampled, where we chose to follow the H20 setup for both surveys following \cite{joudaki/etal:2019}.
Therefore, setup 2 was designed to isolate the effect of moving from the T18 setup to one matching H20. 
We find that setup 1 results in $S_8=0.779^{+0.018}_{-0.038}$, setup 2 shifts $S_8$ to larger values: $0.784^{+0.025}_{-0.024}$ and the outcome of setup 3 which is our fiducial analysis is $S_8=0.755\pm{0.023}$.
We note here that the constraints on $\Om$ and $\sigma_8$ are affected by the choice of priors, which is an interesting topic to be investigated in the future.

Our joint analysis of DES-Y1 and KV450, assuming flat $\Lambda$CDM, results in $S_8=0.755^{+0.019}_{-0.021}$ which is in $3.2\sigma$ tension with the Planck Legacy result (TT,TE,EE+lowE), $S_8=0.834\pm 0.016$. 
\cite{joudaki/etal:2019} drew the same conclusion when combining DES-Y1 and KV450 using $\xi_\pm$, 
$S_8=0.762^{+ 0.025}_{-0.024}$ which is in good agreement with our analysis ($2.5\sigma$ discrepancy with Planck). 
Since we include small angular scales in our analysis we obtain tighter constraints and as such a larger than $3\sigma$ tension with the Planck results. This with the fact that we use COSEBIs instead of 2PCFs in our analysis explains the differences between our results. In particular, the COSEBIs analysis of DES-Y1 data shows a stronger preference for smaller values of $\Om$ compared to the 2PCFs analysis of \cite{joudaki/etal:2019}. 

Planck data are mostly sensitive to early Universe physics, while our cosmic shear analyses probe much lower redshifts ($z \lesssim 1.5$). 
It should be noted that cosmic shear is not the only probe that is in tension with Planck results. 
In fact, currently the largest deviation from Planck results are the $H_0$ measurement of \cite{riess/etal:2019} showing a $4.4\sigma$ discrepancy \citep[see also][for constraints from strong gravitational lensing]{wong/etal:2019}.
There is a large gap between the epoch of recombination and the large scale structures probed by gravitational lensing and the local expansion rate resulting in the measurement of $H_0$. 
The precision of more local probes is increasing, with current cosmic shear surveys gathering more data as this paper is being written. 
As the volume of the data increases the uncertainty on the parameters will decrease. We, therefore, need to make sure that the analysis is truly robust. 

COSEBIs provide reliable statistics for cosmic shear analysis as well as a consistent method for analysing its B-modes, which can point us to systematics in the data. COSEBIs are fairly insensitive to the effect of baryon feedback as they filter out information from small physical scales, allowing for a cosmic shear analysis that is not affected by complicated baryonic physics.
The COSEBIs pipeline, integrated with {\sc CosmoSIS}, is publicly available upon request\footnote{COSEBIs pipeline: ma@roe.ac.uk}.

\begin{acknowledgements}

This project has received funding from the European Union's Horizon 2020 research and innovation programme: We acknowledge support from the European Research Council under grant agreement No.~647112 (CH, MA and CL). TT also acknowledges support under the Marie Sk\l{}odowska-Curie grant agreement No.~797794.
MA acknowledges support from SUPA. CH acknowledges support from the Max Planck Society and the Alexander von Humboldt Foundation in the framework of the Max Planck-Humboldt Research Award endowed by the Federal Ministry of Education and Research. HH is supported by a Heisenberg grant of the Deutsche Forschungsgemeinschaft (Hi 1495/5-1). AC acknowledges support from NASA grant 15-WFIRST15-0008. AK acknowledges support from Vici grant 639.043.512, financed by the Netherlands Organisation for Scientific Research (NWO). KK acknowledges support by the Alexander von Humboldt Foundation.\\

The KV450 results in this paper are based on data products from observations made with ESO Telescopes at the La Silla Paranal Observatory under programme IDs 177.A-3016, 177.A-3017 and 177.A-3018, and on data products produced by Target/OmegaCEN, INAF-OACN, INAF-OAPD and the KiDS production team, on behalf of the KiDS consortium. \\

This project used public archival data from the Dark Energy Survey (DES). Funding for the DES Projects has been provided by the U.S. Department of Energy, the U.S. National Science Foundation, the Ministry of Science and Education of Spain, the Science and Technology Facilities Council of the United Kingdom, the Higher Education Funding Council for England, the National Center for Supercomputing Applications at the University of Illinois at Urbana–Champaign, the Kavli Institute of Cosmological Physics at the University of Chicago, the Center for Cosmology and Astro-Particle Physics at the Ohio State University, the Mitchell Institute for Fundamental Physics and Astronomy at Texas A\&M University, Financiadora de Estudos e Projetos, Funda{\c c}{\~a}o Carlos Chagas Filho de Amparo {\`a} Pesquisa do Estado do Rio de Janeiro, Conselho Nacional de Desenvolvimento Cient{\'i}fico e Tecnol{\'o}gico and the Minist{\'e}rio da Ci{\^e}ncia, Tecnologia e Inova{\c c}{\~a}o, the Deutsche Forschungsgemeinschaft, and the Collaborating Institutions in the Dark Energy Survey.
The Collaborating Institutions are Argonne National Laboratory, the University of California at Santa Cruz, the University of Cambridge, Centro de Investigaciones Energ{\'e}ticas, Medioambientales y Tecnol{\'o}gicas-Madrid, the University of Chicago, University College London, the DES-Brazil Consortium, the University of Edinburgh, the Eidgen{\"o}ssische Technische Hochschule (ETH) Z{\"u}rich,  Fermi National Accelerator Laboratory, the University of Illinois at Urbana-Champaign, the Institut de Ci{\`e}ncies de l'Espai (IEEC/CSIC), the Institut de F{\'i}sica d'Altes Energies, Lawrence Berkeley National Laboratory, the Ludwig-Maximilians Universit{\"a}t M{\"u}nchen and the associated Excellence Cluster Universe, the University of Michigan, the National Optical Astronomy Observatory, the University of Nottingham, The Ohio State University, the OzDES Membership Consortium, the University of Pennsylvania, the University of Portsmouth, SLAC National Accelerator Laboratory, Stanford University, the University of Sussex, and Texas A\&M University.
Based in part on observations at Cerro Tololo Inter-American Observatory, National Optical Astronomy Observatory, which is operated by the Association of Universities for Research in Astronomy (AURA) under a cooperative agreement with the National Science Foundation.\\

{ {\it Author contributions:}  All authors contributed to the development and writing of this paper.  The authorship list is given in three groups:  the lead authors (MA, TT, CH) followed by two alphabetical groups.  The first alphabetical group includes those who are key contributors to both the scientific analysis and the data products.  The second group covers those who have either made a significant contribution to the data products, or to the scientific analysis.}
\end{acknowledgements}


\bibliographystyle{aa}
\bibliography{COSEBIs} 
 

\appendix
\section{Covariance of COSEBIs}

\label{app:covariance}

The covariance of COSEBIs can be written in terms of the sum of four terms,
\begin{equation}
\Cov= \text{SSC}+\text{shape noise}+\text{cosmic variance}+\text{mixed}\;,
\end{equation}
with SSC standing for the super sample covariance which couples the variance of the field within and beyond the survey with each other through disconnected triaspectra and four point functions \citep[see][]{takada_hu:2013}. The shape noise term captures the effect of the intrinsic ellipticity dispersion on the measurements. The cosmic variance term includes a Gaussian and a non-Gaussian term. The Gaussian term encapsulates the covariance associated with a Gaussian shear field, while the non-Gaussian cosmic variance describes any departure from Gaussianity in the shear field resulting in extra information in higher order statistics. Here the relevant terms include the connected convergence trispectra. As \cite{barreira/etal:2018} have shown, in most realistic scenarios, these non-Gaussian components are negligible and therefore we only include the Gaussian term in the calculation of the cosmic variance of COSEBIs. The mixed term corresponds to the cross-correlation between the noise and the Gaussian component of the cosmic variance. 

We can derive the covariance matrix of COSEBIs from their relation to the convergence power spectra (\Eqt\ref{eq:EnBnFourier}) or 2PCFs (\Eqt\ref{eq:COSEBIsReal}). \cite{schneider/etal:2002a} derive the Gaussian covariance for 2PCFs for a general case where they take into account the weights associated with each galaxy, the survey geometry and also the exact number of galaxy pairs. This calculation requires estimating two, three and four-point functions of the galaxy weights for the noise, mixed and cosmic variance terms, respectively. We therefore use the approach in \cite{schneider/etal:2002a} and the COSEBIs relation to 2PCFs for the noise term, but follow the approximation in \cite{joachimi/etal:2008} and the COSEBIs relation to power spectra for the cosmic variance and mixed terms. The covariance in \cite{joachimi/etal:2008} is valid when all angular scales used in the analysis are well within the boundaries of the survey.

The noise term for 2PCFs can be written as,
\begin{align}
\label{eq:CovGauss2PCFs}
C^{ij,kl({\rm noise})}_{\pm\pm}(\theta,\vartheta) &=\langle\Delta \xi_\pm^{ij,\rm noise}(\theta)\; \Delta \xi_\pm^{kl,\rm noise}(\vartheta) \rangle \\ \nonumber
& =\frac{\sigma_{\epsilon, i}^2\: \sigma_{\epsilon,j}^2}{2 N^{ij}_{\rm pair}(\theta)}\delta_{\theta\vartheta}[\delta_{ik}\delta_{jl}+\delta_{il}\delta_{jk}]\;, \\ \nonumber
C^{ij,kl({\rm noise})}_{\mp\pm}(\theta,\vartheta) &=\langle\Delta \xi_\pm^{ij,\rm noise}(\theta)\; \Delta \xi_\mp^{kl,\rm noise}(\vartheta) \rangle =0\;,
\end{align}
where $C^{ij,kl}_{\mu\nu}(\theta,\vartheta)$ is the covariance of 2PCFs for redshift bin pairs $ij$ and $kl$ with $\mu$ and $\nu$ equal to either $+$ or $-$. The weighted number of galaxy pairs, $N_{\rm pair}(\theta)$, can be accurately estimated using the measured number of galaxy pairs in each $\theta$-bin,
 \begin{equation}
  \label{eq:Npair}
 N^{ij}_{\rm pair}(\theta)= \frac{(\sum_{ab} w^i_a w^j_b)^2}{\sum_{ab} (w^i_a)^2 (w^j_b)^2}\;,
 \end{equation}
where the sum goes over all galaxies whose separation $\theta$ falls within the defined angular bin. In \Eqt\eqref{eq:CovGauss2PCFs} 
$\delta_{ij}$ is the Kronecker delta and  $\sigma_{\epsilon, i}$ is the dispersion of the measured galaxy ellipticities in redshift bin $i$, 
\begin{align}
\label{eq:sigmae}
\sigma_\epsilon^2 = & \frac{\sum w^2 \epsilon_1^2}{\sum w^2 (1+m)^2}-\left(\frac{\sum w \epsilon_1}{\sum w (1+m)}\right)^2  \\ \nonumber
+ & \frac{\sum w^2 \epsilon_2^2}{\sum w^2 (1+m)^2}-\left(\frac{\sum w \epsilon_2}{\sum w (1+m)}\right)^2 \;,
\end{align}
where the sum goes over all galaxies in the defined bin and $\epsilon_{1,2}$ are the components of ellipticity. Here we have corrected  the ellipticity dispersion using a multiplicative bias (response in the case of \mcal\!\!, $R=1+m$), because this term is estimated from the data given biased ellipticity measurements and therefore needs to be calibrated similar to the 2PCFs measurement of \Eqt\eqref{eq:xipm_meaure}. Note that the quantity usually quoted as $\sigma_\epsilon$ is the ellipticity dispersion per component of ellipticity and therefore corresponds to $1/\sqrt{2}$ of $\sigma_\epsilon$ as defined in \Eqt\eqref{eq:sigmae}. 

The noise term for the covariance of COSEBIs can be calculated using its relation with the covariance of 2PCFs,
\begin{align}
\label{eq:COSEBICov2PCFs}
C^{ij,kl}_{mn}&=\frac{1}{4}\int_{\theta_{\rm min}}^{\theta_{\rm max}} \d\theta\,\theta\,\int_{\theta_{\rm min}}^{\theta_{\rm max}} \d\vartheta\,\vartheta\\ \nonumber
&\times \sum_{\mu\nu={+,-}} T_{\mu m}(\theta)T_{\nu n}(\vartheta)\; C^{ij,kl}_{\mu\nu}(\theta,\vartheta)\;.
\end{align}
Inserting for $C^{ij,kl}_{\mu\nu}(\theta,\vartheta)$ from \Eqt\eqref{eq:CovGauss2PCFs} to \Eqt\eqref{eq:COSEBICov2PCFs}, we derive,
\begin{align}
\label{eq:NoiseCov}
C^{ij,kl}_{mn}&= \frac{\sigma^2_{\epsilon,i}\sigma^2_{\epsilon,j}}{8} [\delta_{ik}\delta_{jl}+\delta_{il}\delta_{jk}]\int_{\theta_{\rm min}}^{\theta_{\rm max}} \frac{\d\theta\,\theta^2}{n^{ij}_{\rm pair}(\theta)}\\ \nonumber 
&\times [T_{+m}(\theta)T_{+n}(\theta)+T_{-m}(\theta)T_{-n}(\theta)]\;,
\end{align}
where $n^{ij}_{\rm pair}(\theta)\; \d\theta=N^{ij}_{\rm pair}(\theta)$ in the discrete case \citep[see][]{asgari/etal:2019a}. 

The COSEBIs  covariance can also be written in terms of the covariance of the E or B-mode power spectra, $C_{\rm X}^{ij,kl}(\ell,\ell')$, 
\begin{align}
\label{eq:COSEBIsCov}
C_{X(mn)}^{ij,kl} & \equiv  \langle X^{ij}_m X^{kl}_n\rangle-
\langle X^{ij}_m\rangle\langle X^{kl}_n\rangle \nonumber \\
& = \int_0^{\infty} \frac{\d\ell\:\ell}{2\pi} \int_0^{\infty}  \frac{\d\ell'\:\ell'}{2\pi}W_m(\ell)W_n(\ell')\: C^{ij,kl}_{\rm X}(\ell,\ell')\;,
\end{align} 
where X stands for either E or B-mode COSEBIs or power spectra. 
The Gaussian part of the covariance of power spectra is given in \cite{joachimi/etal:2008} and is diagonal, which simplifies \Eqt\eqref{eq:COSEBIsCov} to 
\begin{align}
\label{eq:gaussian}
C_{X(mn)}^{ij,kl}  & = \frac{1}{2 \pi A}\int_0^{\infty} \mathrm{d}\ell\:\ell\:W_m(\ell)W_n(\ell) \\ \nonumber
& \times\left( \bar{P}^{ik}_{\mathrm{X}}(\ell)\bar{P}^{jl}_{\mathrm{X}}(\ell)
+\bar{P}^{il}_{\mathrm{X}}(\ell)\bar{P}^{jk}_{\mathrm{X}}(\ell)\right)\;.
\end{align}
where,
\begin{equation}
\bar{P}^{ik}_{\mathrm{X}}(\ell):= P^{ik}_{\mathrm{X}}(\ell)
+\delta_{ik}\frac{\sigma_{\epsilon,i}^2}{2\bar{n}_i}\;,
\end{equation} 
$A$ is the effective area of the survey and $\bar{n}_i$ is the effective number density of galaxies in redshift bin $i$ which can be calculated with
\begin{equation}
\bar{n}_i= \frac{1}{A}\frac{(\sum_a w_a)^2}{\sum_a (w_a)^2}\;,
\end{equation}
where the sums go over all galaxies within the redshift bin \citep{heymans/etal:2012}.

To calculate the covariance of COSEBIs we use \Eqt\eqref{eq:gaussian} for the Gaussian component of the cosmic variance and the mixed term. We remove the terms that only depend on shape noise from \Eqt\eqref{eq:gaussian} and instead use \Eqt\eqref{eq:NoiseCov} for the noise only terms. 
The super sample term is included by integrating over the super sample covariance of the convergence power spectrum using \Eqt\eqref{eq:COSEBIsCov}.

In addition to these terms, the uncertainty in the multiplicative shear bias, $m$, in \Eqt\eqref{eq:xipm_meaure} can be absorbed in the covariance matrix through its expected dispersion, $\sigma_m$. This additional covariance term is
\begin{equation}
\label{eq:sigmam}
C^{ij, kl (m-{\rm bias})}_{nm}=4\sigma_m^2 E_n^{ij} E_m^{kl}\;.
\end{equation}
The value of $\sigma_m$ can be estimated using image simulations and can in principle depend on the redshift bin \citep{hildebrandt/etal:2017}. Alternatively, $m$ can be set as a free parameter in the analysis with a Gaussian prior with a variance of $\sigma_m^2$ as was done in \cite{troxel/etal:2018a}. These two approaches are equivalent, however, including the $\sigma_m$ through the covariance decreases the number of parameters that need to be sampled and therefore is our preferred approach. In our analysis of the DES-Y1 data we keep the covariance matrix constant over the different setups and only include the $\sigma_m$ term when $m$-bias is not allowed to be a free parameters in the analysis.

\section{Supplementary plots}
\label{app:triangle}

\begin{figure}
   \begin{center}
     \begin{tabular}{c}
      \includegraphics[width=\hsize]{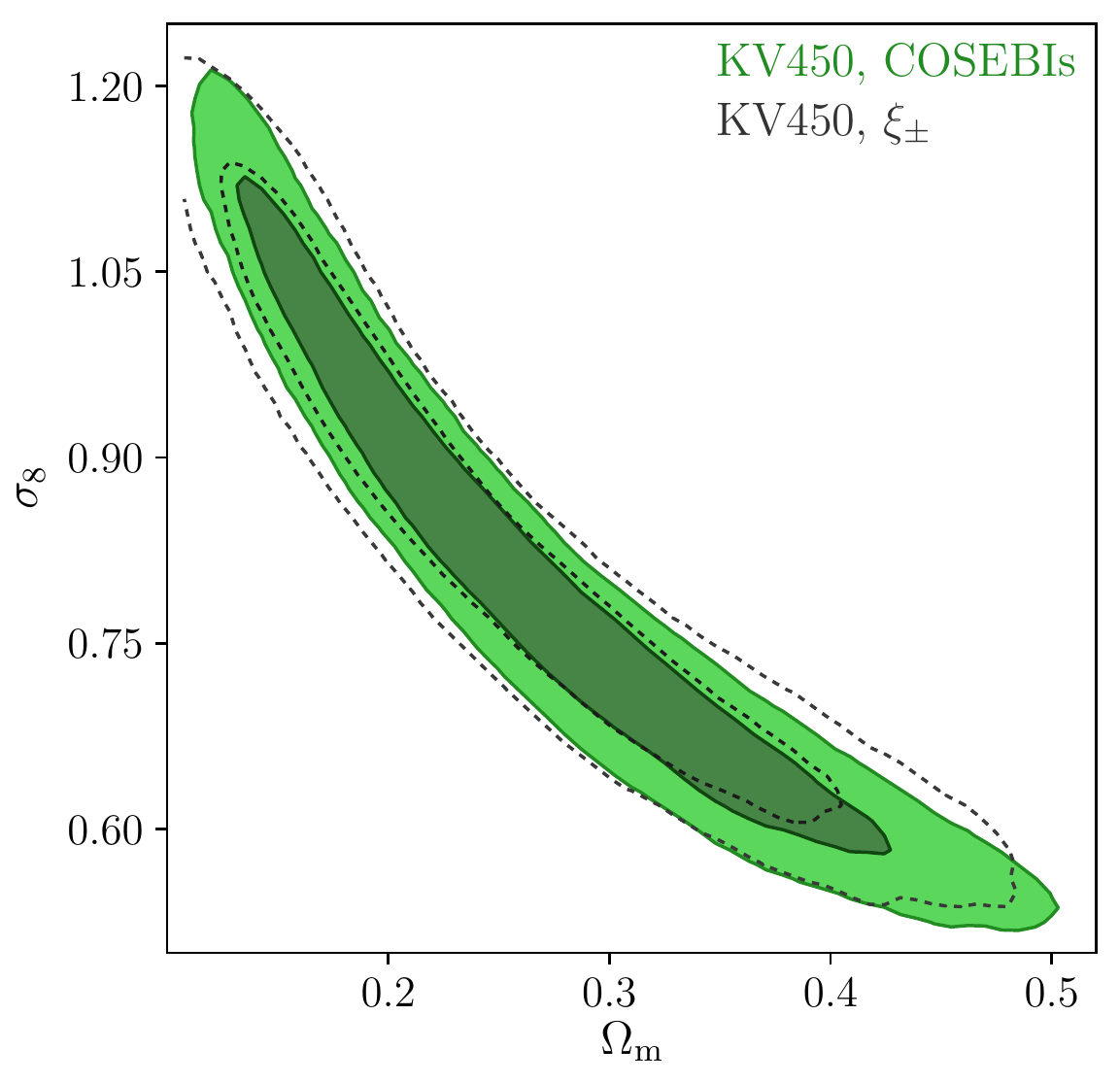}
     \end{tabular}
   \end{center}
     \caption{\small{COSEBIs (green) versus 2PCFs (dashed) contours for KV450. The constraints are shown for $\sigma_8$ and $\Om$. }}
     \label{fig:KV450_banana}
 \end{figure}

\fig\ref{fig:KV450_banana} shows a comparison between COSEBIs and 2PCFs constrains for KV450. The results are shown in $\sigma_8$ and $\Om$. COSEBIs (green) and 2PCFs (dashed) contours show very similar results. 

\begin{figure*}
   \begin{center}
     \begin{tabular}{c}
      \includegraphics[width=\hsize]{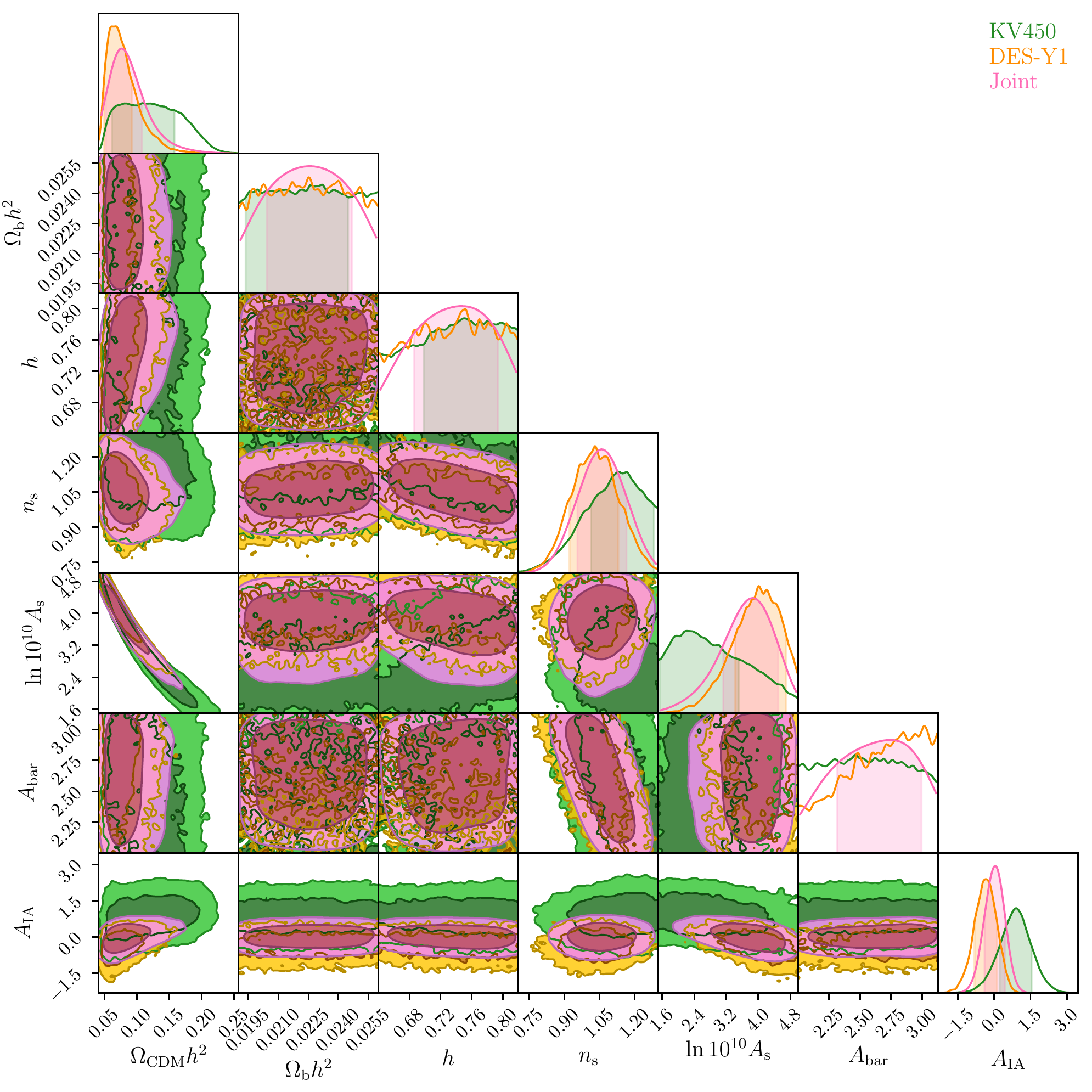}
     \end{tabular}
   \end{center}
     \caption{\small{Constraints on cosmological parameters: $\Omega_{\rm CDM}h^2$, $\Omega_{\rm b} h^2$, $h$, $n_{\rm s}$, $\ln( 10^{10} A_{\rm s})$ and astrophysical parameters: $A_{\rm bar}$ and $A_{\rm IA}$. Joint constraints are shown in pink, KV450 results in green and DES-Y1 in orange (see \tab\ref{tab:setups}).}}
     \label{fig:triangle}
 \end{figure*} 

\fig\ref{fig:triangle} shows the constraint on the cosmological and astrophysical parameters obtained from KV450, DES-Y1 and their joint analysis. We use the H20 setup in \tab\ref{tab:setups} and DIR calibrated redshift distributions for all three cases.
Here we see that our joint analysis can put meaningful constraints on not only $\ln(10^{10} A_{\rm s})$ and $\Omega_{\rm CDM}h^2$, but also on $n_{\rm s}$ and $A_{\rm IA}$. 
Our $n_{\rm s}$ values are fully consistent with \cite{Planck2018} results. In addition, our joint constraints on $A_{\rm IA}=0.05^{+0.40}_{-0.44}$, are consistent with the direct measurements of \cite{johnston/etal:2019}, who reported $A_{\rm IA}=1.06^{+0.47}_{-0.46}$ with a $1.6\sigma$ difference. The corresponding values for KV450 are $A_{\rm IA}=0.92^{+0.63}_{-0.69}$ which shows a  $0.2\sigma$ difference with \cite{johnston/etal:2019} and the DES-Y1 results are $A_{\rm IA}=-0.35^{+0.47}_{-0.46}$ with a very mild, $2.1\sigma$, difference with the \cite{johnston/etal:2019} results.
Here we have used a joint model with a single free parameter for the intrinsic alignment of galaxies. The effect of using separate parameters for each survey, as well as assuming redshift evolution for the intrinsic alignment signal has been explored in \cite{joudaki/etal:2019}, where they find that these alternatives have little impact on the cosmological constraints.


\end{document}